\documentclass[a4paper,useAMS,usenatbib]{mn2e}
\newcommand{\captA}{These are logarithmic false colour grayscale
maps of the number density. Columns correspond to different simulations and 
time increases downwards. Arrows indicate velocity. See Table~1 for details on the model
parameters.}

\newcommand{\one}{Model~1}
\newcommand{\two}{Model~2}
\newcommand{\twoc}{Model~2b}
\newcommand{\three}{Model~3}
\newcommand{\four}{Model~4}
\newcommand{\five}{Model~5}
\newcommand{\six}{Model~6}
\newcommand{\seven}{Model~7}
\newcommand{\eight}{Model~8}

\usepackage{graphicx}
\usepackage{subfigure}

\title[Simulating the Morphological Evolution of PN]
{From Bipolar to Elliptical: Simulating the Morphological Evolution 
of Planetary Nebulae}
\author[M. Huarte-Espinosa et al.]
{\parbox{\textwidth}{M. Huarte-Espinosa,$^{1}$\thanks{E-mail: \textit{martinhe@pas.rochester.edu}}
A.~Frank,$^{1}$ 
B.~Balick,$^{2}$
E.~G.~Blackman,$^{1}$
O.~De~Marco,$^{3,4}$
J.~H. Kastner$^{5}$ and 
R.~Sahai$^{6}$}\vspace{0.4cm}\\
\parbox{\textwidth}{$^1$Department of Physics and Astronomy, University of Rochester, 600 Wilson Boulevard,
Rochester, NY, 14627-0171 \\
$^2$Department of Astronomy, University of Washington, Seattle, WA 98195 \\
$^3$Department of Astrophysics, American Museum of 
Natural History, Central Park West at 79th Street, New York, NY 10024 \\
$^4$Department of Physics, Macquarie University, Sydney NSW 2109, Australia \\
$^5$Rochester Institute of Technology, 54 Lomb Memorial Drive, Rochester, NY 14623, USA \\
$^6$NASA/JPL, 4800 Oak Grove Drive, Pasadena, CA 1109, USA}}
\begin{document}

\date{Received \today}
\pagerange{\pageref{firstpage}--\pageref{lastpage}} \pubyear{2012}
\maketitle
\label{firstpage}

\begin{abstract}
In this paper we model the evolution of PrePlanetary Nebula (PPN)
and Planetary Nebula (PN) morphologies as a function of nebular
age. The aim of the work is to understand if shape transitions from
one evolutionary phase to the other can be driven by changes in
the parameters of the mass-loss from the central star.
We carry out 2.5D~hydrodynamical simulations of mass-loss at the
end stages of stellar evolution for intermediate mass stars.  Changes
in wind velocity, mass-loss rate and mass-loss geometry are tracked.
We focus on the transition from mass-loss dominated by a short
duration jet flow (driven during the PPN phase) to mass-loss driven
by a spherical fast wind (produced by the central star of the PN).
% 
%%% sahai '12 
   Our results show that 
	%including a period of jet formation in the 
   %temporal sequence of PPN to PN can produce a transition from bipolar 
   %outflows to elliptical nebula.   
   % 
   %In particular  we find that  
%%% 
% 
while jet driven nebulae can be expected to be dominated by bipolar
morphologies, systems that begin with a jet but are followed by a
spherical fast wind will evolve into elliptical objects. 

Systems
that begin with an aspherical AGB wind evolve into butterfly shaped
nebula with, or without, a jet phase. In addition, our models show
that spherical nebulae are highly unlikely to derive from either
bipolar PPN or elliptical PN over relevant time scales.  The morphological transitions seen
in our simulations may however provide insight into the driving mechanisms
of both PPN and PN as point to evolutionary changes in the
central engine.
\end{abstract}


\section{INTRODUCTION}

Planetary Nebulae (PN) 
expand at a~few~10\,km\,s$^{-1}$ away from evolved intermediate-mass
stars (with 
main sequence progenitor masses $\sim\,$1$-$8\,M$_{\odot}$). These nebulae exhibit a rich
variety of morphologies with the main categories being spherical,
elliptical and bipolar
\citep[for a review see][]{balick02}. %BB%
%BB%Additional classifications including point symmetric,
%BB%quadrupolar and irregular morphologies
%BB%%from garcia segura apj, 517, 767 
%BB%(Chu, Jacoby, \& Arendt 1987; Schwarz, Corradi \& Melnick 1992;
%BB%Stanghellini, Corradi, \& Schwarz 1993; Manchado, Stanghellini \&
%BB%Guerrero (1996); \citealp{manchado96b}).  
Spherical PN 
%orsola
   %have found explanation through
   have been explained by 
time-dependent interacting stellar wind models
(ISW;  Kwok, Purton, \& Fitzgerald 1978) that describe the nebula
as the collision of a fast ($\sim\,$1000\,km\,s$^{-1}$) tenuous
% 
%%% orsola 
   %($\sim\/$10$^{-7}
	(between about 10$^{-8}$ and 10$^{-7}$\,
%%% 
% 
M$_{\odot}$\,yr$^{-1}$) stellar wind driven by
the hot central star 
of PN %orsola 
(CSPN) with a dusty slow ($\sim\,$10\,km\,s$^{-1}$)
dense ($\sim\/$10$^{-4}$\,M$_{\odot}$\,yr$^{-1}$) wind, or circumstellar envelope, 
ejected during the earlier AGB phase.
The nebulae are thus formed when the fast wind catches up with the AGB wind, 
driving a shocked shell through it.

The formation of bipolar PN and jets has been addressed by generalized ISW models (GISM) in which the 
ejected AGB
envelope is characterized by either a toroidal density distribution
or an aspherical velocity field
%GGS
%
%joel
   %~(Kwok, Purton, \& Fitzgerald 1978; 
%%%
%
~(Kahn \& West 1985;
\citealp{balick87,icke88,huggins07}; Frank \& Mellema~1994;
\citealp{icke88}; Icke, Preston, \& Balick, 1989; Mellema, Eulderink,
\& Icke, 1991; Icke, Balick, \& Frank, 1992; \citealp{frank94};
Dwarkadas, Chevalier, \& Blondin, 1996). In some of 
these models, asymmetries in AGB envelopes
develop as a result of a binary stellar system, in which
an AGB star interacts with a companion
\citep{arredondo,edgar}, or an AGB star and a companion
share a common envelope evolution (\citealp{iben91,livio93,soker97};
\citealp{sandquist98}; 
\citealp[and references therein]{soker98}; 
\citealp{demarco03,nordhaus07,passy12}). 
Note, however, that AGB envelopes generally show spherical %rarely show aspherical %JOEL 
shapes (Bujarrabal \& Alcolea 1991; Kahane \& Jura 1994; \citealp{stanek95,groeneweger96}).  
Some AGB observations show seeds of asymmetry.

The original GISW 
model 
%explored by Balick, Icke and others %ORSOLA 
(see e.g. Balick \& Frank 2002)
assumed a spherical fast wind driven from the CSPN  
%(Miszalski et al. 2009a) ORSOLA
expanding into an aspherical AGB wind. While these models
were able to 
generate hot, low density jet flows via shock focusing at the inner shock,
they were less successful at recovering dense colder jets like
those seen in YSOs.  Since many PPN and young PN show evidence for
such narrow jets, Sahai \& Trauger (1998) suggested that collimated
PN flows created at, or near, the central engine were the real
drivers of early PN morphology.  
%%%JOEL
   Such jet formation has been proposed as a natural 
	consequence of interacting binary evolution
   (\citealp[][2004]{lee03}; \citealp{akashi08,lee09}).
   \citet{lee03} %http://adsabs.harvard.edu/abs/2003ApJ...586..319L 
	carried out numerical simulations and synthetic
	observations of a jet (or collimated fast wind) interacting 
	with a spherical AGB wind. They found that both the dynamics 
	and emission of PPN, and young PN, shells depend on the velocity 
	and geometry of the jet. %Their model also reproduced the main morphological 
	%features observed in the W1~lobe of CRL~618; a highly collimated
	%lobe with a bowlike emission structure at its tip.
		%difficulty producing the bright emission structures seen
		%along the body of the lobe.  temperatures at the tip of the
		%lobe in our simulations are higher than observed.  The
		%collimated fast wind in CRL 618 is unlikely to be steady
		%and is not radiatively driven.
   %
   Using a similar approach
	\citet{akashi08} %http://adsabs.harvard.edu/abs/2008MNRAS.391.1063A 
   modeled the effect of short-duration jets expanding into
	a spherical, slow AGB wind on the observed shape of evolved PN.
	The simulations of \citet{akashi08} covered timescales about 
	an order of magnitude longer than the ones of \citet{lee03}. 
	\citet{akashi08}
	found that the AGB mass-loss history also has a profound influence
	on the nebular structure, and that the head of the lobes 
	move sufficiently fast to excite some visible emission lines
	(see \citealp{lee03} for detailed results of 
	optical forbidden line emission which is excited at the heads of the 
	lobes during the PPN phase).
	%Lee n sahai 2004
	   %http://adsabs.harvard.edu/abs/2004ApJ...606..483L
	%Lee n sahai 2009
	   %http://adsabs.harvard.edu/abs/2009ApJ...696.L
%%%
%
%%% joel feb '12
   These studies did not address the question as to how the older,
   AGB spherical outflows would respond to ``fresh'', collimated
   outflows from the PN central engine.

%Adam, jan12
  In this paper we focus on a theoretical exploration of the role
  that changing mass-loss geometries from the central engine can
  play in driving changes in nebular morphology.  If PPN drive
  collimated outflows off the central engine but PN drive primarily
  spherical outflows then how will the nebula respond as it expands?
  The aim is to understand how nebulae
  change form as their hydrodynamic driving changes.
  Such process is of interest
  because it might help to understand the history of the central
  engine which may be a single star, a recently processed common
  envelope system or a binary with an accretion disk. Thus there
  is the hope that one might be able to use morphological clues in
  an individual nebula to recover something of the history of the
  central object in terms of its mass-loss and the conditions which
  govern it.
%%%

We note that there is a broader observational motivation for this
study.  The classifications of PPN and PN by morphological type
appear to show significant statistical differences. The general
trends indicate that (i) around~70 to~90\% of PPN and young PN are
bipolar or more complex \citep{sahai98,sahai11}, (ii) about 20\%
of mature PN are round, 70\% are elliptical and about~10\% are
bipolar (see \citealp{miszalski09a,parker06}; \citealp{lagadec11};
the Planetary Nebula Image Catalog of Bruce
Balick\footnote{http://www.astro.washington.edu/users/balick/PNIC/} and
references therein).  There are however considerable uncertainties
in these classifications, e.g. (i) telescope resolution limitations;
(ii) stellar population selection effects; (iii) differences in
nebular class definitions (for example ring-like nebulae that are
actually toroids included in spherical morphological classes); (iv)
the well-established correlation between PN morphology and the
progenitors' mass, i.e., bipolar and ring-like PN generally found
at lower galactic scale heights, characteristic of higher-mass
progenitors \citep[see e.g.][]{corradi95,kastner96,manchado11}.
These factors, along with the fact that most attempts to classify
PN by morphological type do not consider the ages of the nebulae
(but see \citealp{sahai98,sahai11} for specific studies on young
PN), complicate the interpretation of temporal changes in nebular
morphology.

%
%%%joel feb'12
   %
	%%%adam feb'12
     %An additional problem in the context of nebular shape classification
     %arises when comparing the morphological distributions of PPN and
     %young PN with those of older ones. The fraction of spherical PN is
     %virtually zero among the former, and increases considerably in the
     %latter. Deep searches for PN which detected many faint objects (e.g.
     %the MASH survey of \citealp{parker06} and \citealp{miszalski08})
     %show the overall round PN fraction to increase from a few~to
     %$\sim\,$20\%. There are few young counterparts to these spherical,
     %rimmed PN \citep[e.g. Abell~39, IC~418][] {jacoby00}.

     %There remains an unresolved issue concerning the statistical inconsistency between the morphologies
     %of PPN and PN then. If the imbalance between PPN and  PN statistics holds up under further scrutiny then it implies that some objects may undergo a shape transformation as they expand.  In particular,
     %one can imagine a scenario in which bipolar PPN expand and become
     %elliptical or rounder PN. Such process could be caused by different
     %nebular driving mechanisms during the early and late nebular
     %evolutionary phases.

   %%%
	%

  If the imbalance between PPN and PN statistics holds up under further scrutiny 
  then it implies that some objects may undergo a shape transformation as they expand. 
  In particular, one can imagine a scenario in which bipolar PPN hydrodynamically transform into 
  elliptical or rounder PN. Such a transformation could be caused by different 
  nebular driving mechanisms during the early and late nebular evolutionary phases.
  % 
  %%% sahai feb '12
    %We note the transformation also depends on the 
    %relative brightness of the lobes compared to the waists
    %in PPN which show waists \citet[see][]{sahai11}, but this is beyond
    %the scope of this study.
	 %
	 %  OR
	 %
    %We note there is an additional possibility, one which has
    %been studied observationally by \citet{sahai11} and is relevant
    %to the evolution of the waist region of PPN which show waists,
    %as they evolve into the PN phase. \citet{sahai11} have found
    %that the apparent shape of these nebulae depends not only on
    %the lobes' spatial distribution and structure, but also on the
    %relative brightness of the lobes compared to the waists,
    %particularly when the lobes have become too tenuous to be seen.
    %These effects will be considered in a future work.
  %%% 
  % 

  Irrespective of issues of observational statistics, in this paper
  we take on the issue of morphological changes in PPN/PN from a
  theoretical perspective. Starting from a set of basic physical
  ingredients and using simulations, we aim  to understand how shape
  transitions might occur during the evolution of an individual
  post-AGB system and how such morphological transitions  depend
  on the parameters of the mass-loss process of the central star.
  The morphological evolution of single sources should be studied
  from first principles to better inform the interpretation of
  observations.
  %In our investigation of the potential for profound morphological changes during 
  %PPN/PN evolution, 
  Thus
%%% 
% 
we consider the influence of an initial collimated jet phase
during the PPN and its effect on the previously deposited circumstellar
AGB wind. The jet phase is followed by a classical fast wind with
a spherical geometry.  Different distributions of the AGB wind's
velocity and density are explored as are other parameters for the
jet/wind interactions.  Our choices for initial and boundary
conditions are based on a reasonable balance between observational
expectations and computational expediency.  
We note \citet{soker90}
has studied PN ansae using a similar wind/jet sequence, but
here we carry out a more extensive study and explore a range of 
parameters across both PPN and PN evolution.
 
This paper is organized as follows: in section~\ref{model} we
describe the methodology and numerical code used in the study, as
well as our model and implementation of the AGB wind and post-AGB
stellar outflows. The results of the simulations are presented in
section~\ref{results}. In section~\ref{discu} we compare our results
with the general trends found by observations by means of synthetic
emission maps and discuss how the results bear on issues of PN
shaping. Conclusions are presented in section~\ref{conclu}.

\section{MODEL AND METHODOLOGY}
\label{model}

We carry out a set of Eulerian-grid numerical simulations to follow
the interaction of collimated jets and winds. The initial and inflow
conditions are meant to model nebular evolution from the AGB
to the mature PN phases. Wind dynamics are modeled and followed in
2.5 dimensions (cylindrical symmetry) using the equations of radiation
hydrodynamics. The effects of optically thin cooling have been
included using the cooling tables of \citet{dm}.  We note that
radiative cooling will play an important role in the shocks driven
by the jet and also in the ``rim'' of swept up AGB material driven
by the fast wind. Because of its high velocity the shocked fast
wind should form a high temperature ``hot bubble'', $T_{\rm hb} \sim
10^7$\,K, that will not cool effectively during the evolution of
the PN.  Observations have established, however, that only certain
PN show such hot bubble emission and that the temperature is usually
lower, by factors of order 10 to 100, than that predicted by the
jump conditions \citep[e.g.][and references therein]{kastner08}.

The hydrodynamic equations are solved with the adaptive mesh
refinement (AMR) numerical code AstroBEAR1.0\footnote{
http://bearclaw.pas.rochester.edu/trac/astrobear/wiki}.  
In particular the Euler
equations with cooling source terms are solved using a second-order
MUSCL-Hancock shock capturing scheme and Marquina flux functions
\citep{abear}. While AstroBEAR1.0 is able to solve the equations of
magnetohydrodynamics (MHD) and to compute several microphysical
processes, such as gas self-gravity and heat conduction, we do not
consider these processes in the present study.  
%BASE ON ORSOLA'S COMMENTS:
  %While the jet is likely
  %formed via MHD processes at the central engine but the fields are
  %not likely to dominate global morphologies the nebular scales
  %\citep{hartigan07}.
  %
  %%%JOEL 
     We justify these omissions at present on the basis that
	  %Although 
  %%% 
  % 
  jets are likely formed via MHD processes at the central engine, but
  magnetic fields are not likely to dominate global morphologies at nebular 
  scales \citep{hartigan07}.
%%%

The numerical domain of the simulations is $1$\,pc on each side.
We use cylindrical coordinates $(r,z)$ in a box that includes
explicit calculations of the pole to avoid numerical issues
there\footnote{
  We moved the bottom grid boundary away form the nebular pole. Cells along 
  this axis were evolved as grid cells, not as boundary (ghost) ones, by solving the equations
  of hydrodynamics. Hence we do not introduce any known numerical artifacts related to 
  reflective boundaries on those cells.
}.
Thus our simulation extends from $-r_{max}$ to $r_{max}$,
where $r_{max}=\,$0.5pc.
Extrapolation, or outflow, boundary conditions are used at $r = \pm\,$0.5\,pc
(above and below the $z$ axis) and at $z=\,$1\,pc.
Reflective boundary conditions are 
imposed %used %JOEL 
at the ``equator'' of the
system ($z=0$).  We consider cells with radii $R < r_w$, where
\hbox{$r_w\sim\,$6000\,AU}, to be the ``nozzle'' of the jets/wind and
we apply %JOEL 
inflow boundary conditions there (see below).

A coarse grid of 128$\times$128\,cells is employed along with two 
levels of AMR refinement, each increasing the resolution by a factor of 2.  
Thus the simulations attain a maximum resolution of
\hbox{512$\times$512} with a $\Delta X_{min} \sim\,$403\,AU. Typical
simulation flow times are of order~13000\,yr which 
% 
%%% ORSOLA APR2011 
   %makes our work some of the longest 
   is about ten times longer than previous simulations of PN evolution 
	(\citealp[e.g. compare with][2004]{lee03}; \citealp{lee09,akashi08}). 
%%%
%
We use BlueHive\footnote{
https://www.rochester.edu/its/web/wiki/crc/index.php/ BlueHive\_Cluster},
an IBM parallel cluster maintained at the 
Center for Integrated Research Computing of the
University of Rochester, to run simulations for about 20\,hrs, using
16~processors.

%%%%%%%%%%%%%%%%%%%%%%%%%%%%%%%%%%%%%%%%%%%%%%%%%
\subsection{Outflow Phases}
\label{epis}

Three outflow episodes are considered in the simulations.  The first
   one, the ``AGB wind'', %Balick
constitutes AGB initial conditions (e.g. \citealp{knapp85})
set on the 
entire %BASED ON ORSOLA'S COMMENTS 
computational grid. 
% 
%%%% ORSOLA+MARTIN 
   These conditions are such that %By assumption 
%%% 
% 
the edge of the AGB envelope is located outside our computational domain. 
Our AGB wind/circumstellar environment has an isotropic 
radial velocity field of $v_{AGB}=\,$10\,km\,s$^{-1}$ and 
a mass-loss rate of
\hbox{$\dot{M}_{AGB}=\,$1$\times$10$^{-5}$\,M$_{\odot}$\,yr$^{-1}$}.
These values are representative of observed AGB wind properties
(Hrivnak et al. 1989; Bujarrabal et al. 2001).
%
%%JOEL APR 2011
   The AGB wind is set up with a uniform temperature of 500\,K
	which best represents the inner region of the envelope. We 
	would actually expect a temperature gradient, with the outer temperature being
	colder, however this simplification does not affect the current model.
%%%

The second outflow phase, the ``jet'', begins immediately upon the
initiation of the simulation. A collimated jet, with an opening
angle of zero~degrees, is injected at cells with radii~$< r_w$ ($\sim\,$6000\,AU).
The jet's velocity, temperature and mass-loss rate are
200\,km\,s$^{-1}$, 500\,K and 
\hbox{$\dot{M}_{j}=\,$5$\times$10$^{-6}$\,M$_{\odot}$\,yr$^{-1}$}
($=$ 0.5
$\dot{M}_{AGB}$), respectively.  The jet is active for
approximately 100\,yr which is about 1\% of the average total
simulation run time. The launch and collimation of our jets
are assumed to occur in the ``central engine'' (likely a binary or
strongly magnetized stellar winds\footnote{
% 
%%%ORSOLA
   We do not delve here on the actual physical causes of jets and
   other phenomena. We merely study the effect of sequences of
   plausible events on the PN morphology.})
%%% 
% 
located at sub-resolution scales
of order~10\,AU. The jet injection timescale is representative of
expansion times (which are of order 500\,yr) of some PPN and PN
which can be directly imaged \citep[e.g. see the PPN jet sample
in][and references therein]{huggins07}.
   %%%Our jet injection timescale of approximately~100\,yr is 
   %%%representative of observed collimated PPN expansion times 
   %%%%\citep[$t_{exp} \sim 500\,yr$,
   %%%\citep[which are of order 500\,yr, see][and references therein]{huggins07}.  
   %%%Jet opening angles of $\la\,$30$^{\circ}$ %zero~degrees %%SAHAI
   %%%are consistent with observations of both PPN and PN for which bipolar
   %%%lobes can be directly imaged 
   %%%%
   %%%%AF: THESE ARE THE WRONG REFS FOR THIS 
   %%%%\citealp{schwarz92,manchado96b}; Sahai \& Trauger 1998; \citealp{gorny99}).
   %%%\citep[e.g. see the PPN jet sample in][and references therein]{huggins07}.
Jet mass-loss rates are not well characterized so we have used an
intermediate value (yet higher than those used in previous
simulations, e.g. see \citealp{lee09}, and references therein)
between the AGB and CSPN
   evolutionary stages, %joel may2011
consistent with 
both 
momentum excesses observed in PPN flows \citep{bujarrabal01}
and the observations of A~63 carried out by Mitchell et al. (2007).

The third outflow phase
we consider, 
   the ``fast wind'', %balick
is a standard central stellar fast wind \citep{kwok78} with a
spherical velocity field continuously injected into the computational
domain through the nozzle. In our three-phase models the spherical
wind remains active throughout the remainder of the
simulation 
after the jet is turned off. %JOEL 
The fast wind has a mass-loss rate that decreases in time over
$\sim\,$10$^4$\,yr from 5$\times$10$^{-7}$~to
\hbox{5$\times$10$^{-9}$\,M$_{\odot}$\,yr$^{-1}$}, following the
results of \citet[and references therein]{perinotto04}.  Once
$\dot{M}$ reaches \hbox{5$\times$10$^{-9}$\,M$_{\odot}$\,yr$^{-1}$}
it remains constant at that rate until the end of the simulation.
Simultaneously, the fast wind increases in speed from
200~to 2000\,km\,s$^{-1}$ in a manner that conserves wind ram
pressure. 
This set of final fast wind conditions was chosen for computational 
expediency. Gas temperature in the wind at the nozzle is set such
that the wind maintains a constant Mach number of~20
(again, for computational expediency; the model results 
are not sensitive to this).

We now review the suite of simulations carried out for the study.
To isolate the effect of the jet we ran a series of
two-phase simulations in which only the jet is launched into the
AGB wind (Models~1 and~4, see Table~1). After the jet episode ends
in these models no further gas is injected into the grid.  Also, the
simulations represented by 
Models~2, 2$^*$, 2b, 2b$^*$, 5 and~7 %\two, \two*, \twoc, \twoc*, \five\ and \seven\ 
are two-phase models in which the fast wind is injected
directly into the AGB wind.  The simulations represented by 
Models~3, 3$^*$ and~6 %\three, \three* and \six\ 
are three-episode scenarios in which the fast
wind is injected immediately after the jet, at time $t \sim 100\,yr$
(see Table~1).  The simulation represented by \eight\ is almost the
same as \three, except that the former has a wind-quiescent interlude
that lasts~$\sim\,$400\,yr between the jet and the fast wind episodes.
The interlude represents an arbitrary timescale  of order~4 times
the jet's duty cycle (or active time) chosen to be long enough to allow the jet
driven cavity to evolve but short enough to be consistent with
PPN/CSPN transition timescales.

\begin{figure*}
\centering
  \includegraphics[width=.495\textwidth,bb=0.75in 1.00in 5.75in 3.9in,clip=]{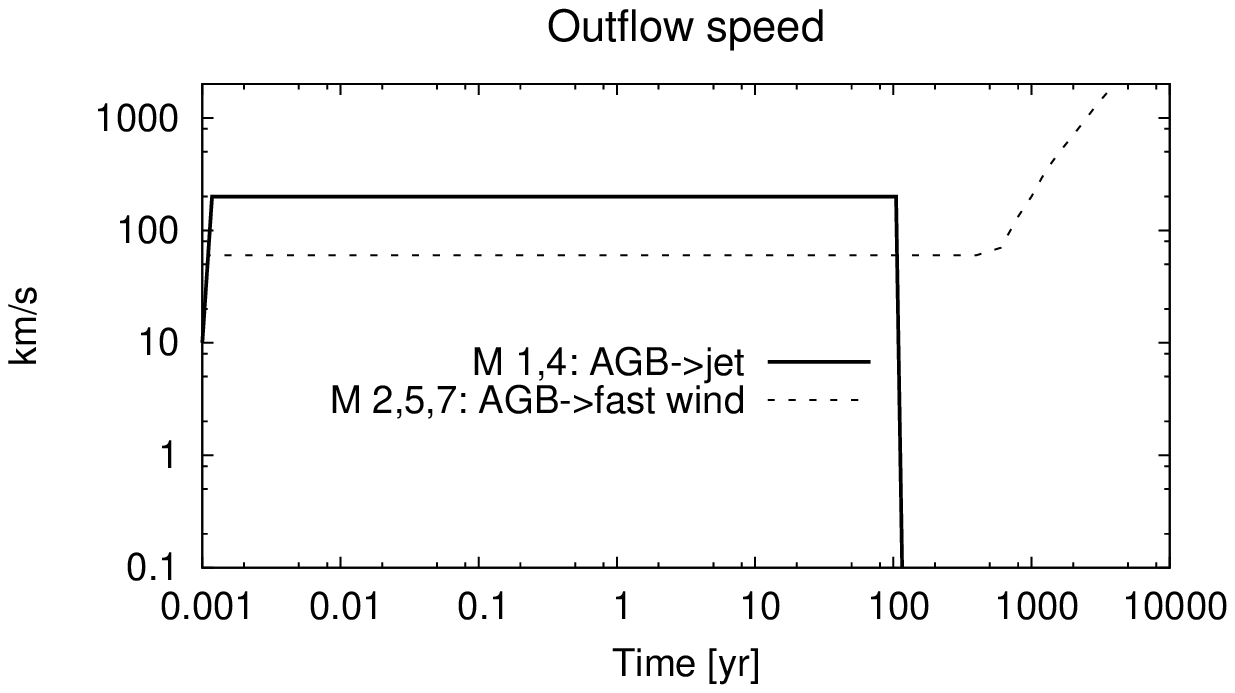}
  \includegraphics[width=.495\textwidth,bb=0.75in 1.00in 5.75in 3.9in,clip=]{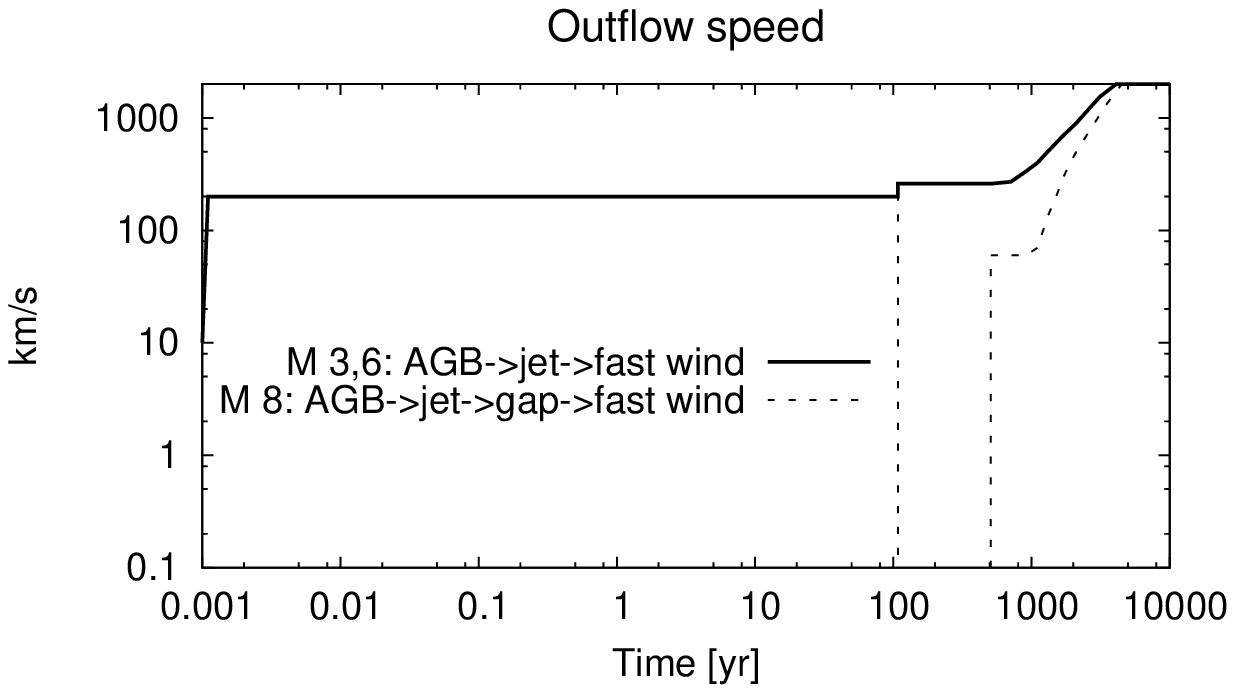}\\
  \vskip.5cm                                 
  \includegraphics[width=.495\textwidth,bb=0.75in 1.00in 5.75in 3.9in,clip=]{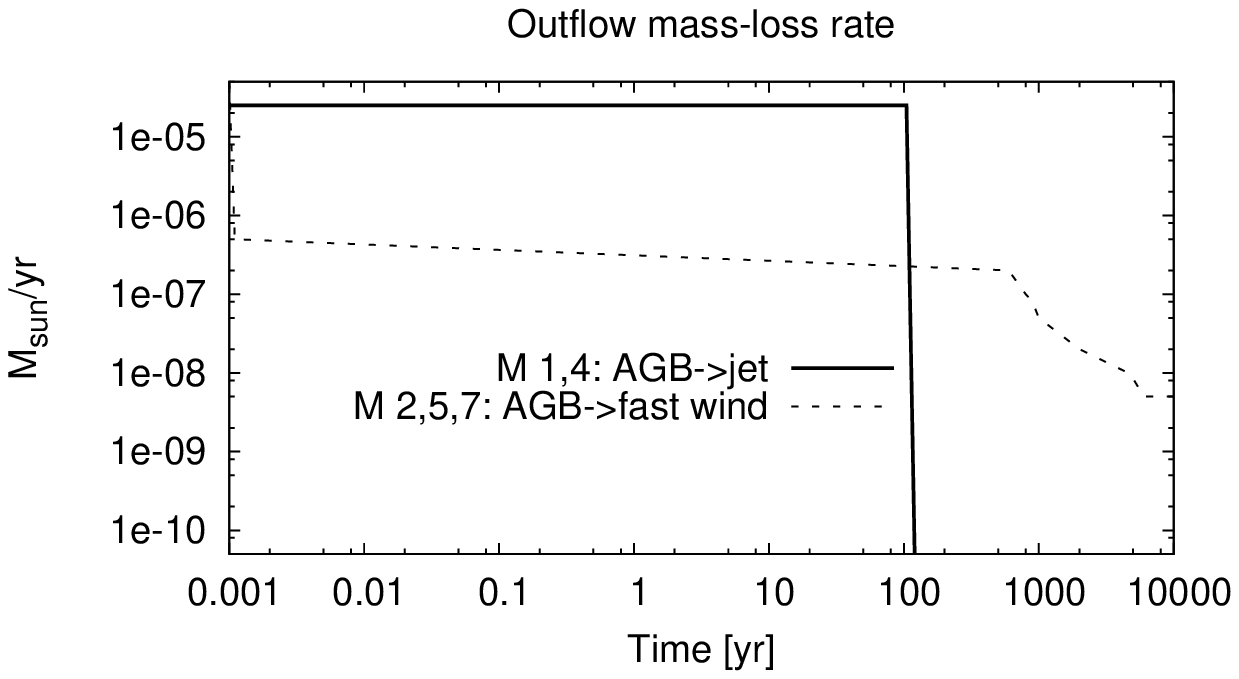}
  \includegraphics[width=.495\textwidth,bb=0.75in 1.00in 5.75in 3.9in,clip=]{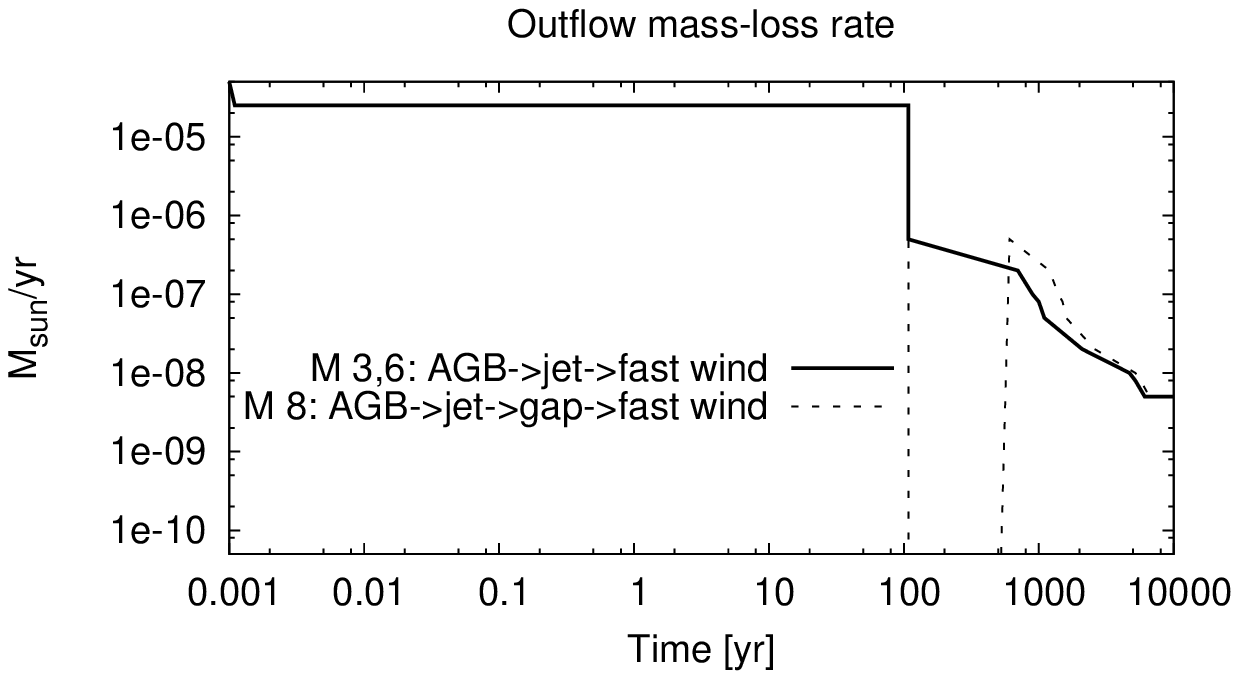}\\
  \vskip.5cm                                 
  \includegraphics[width=.495\textwidth,bb=0.75in 0.80in 5.75in 3.9in,clip=]{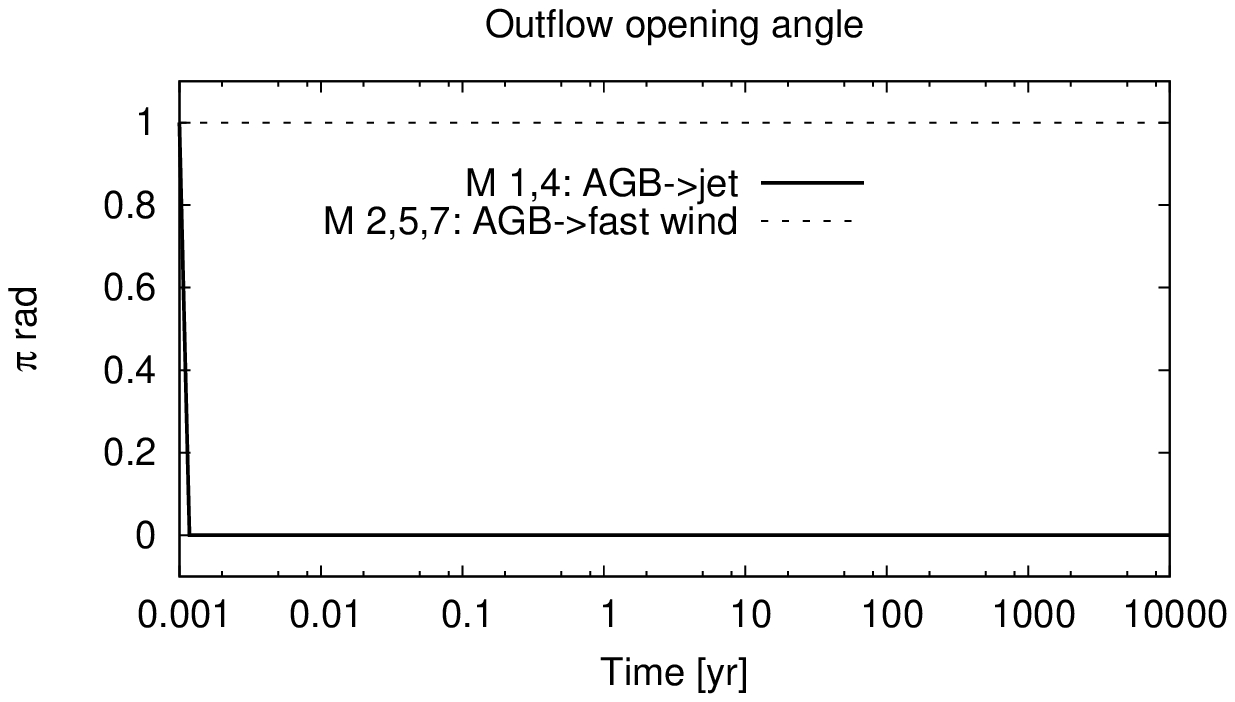}
  \includegraphics[width=.495\textwidth,bb=0.75in 0.80in 5.75in 3.9in,clip=]{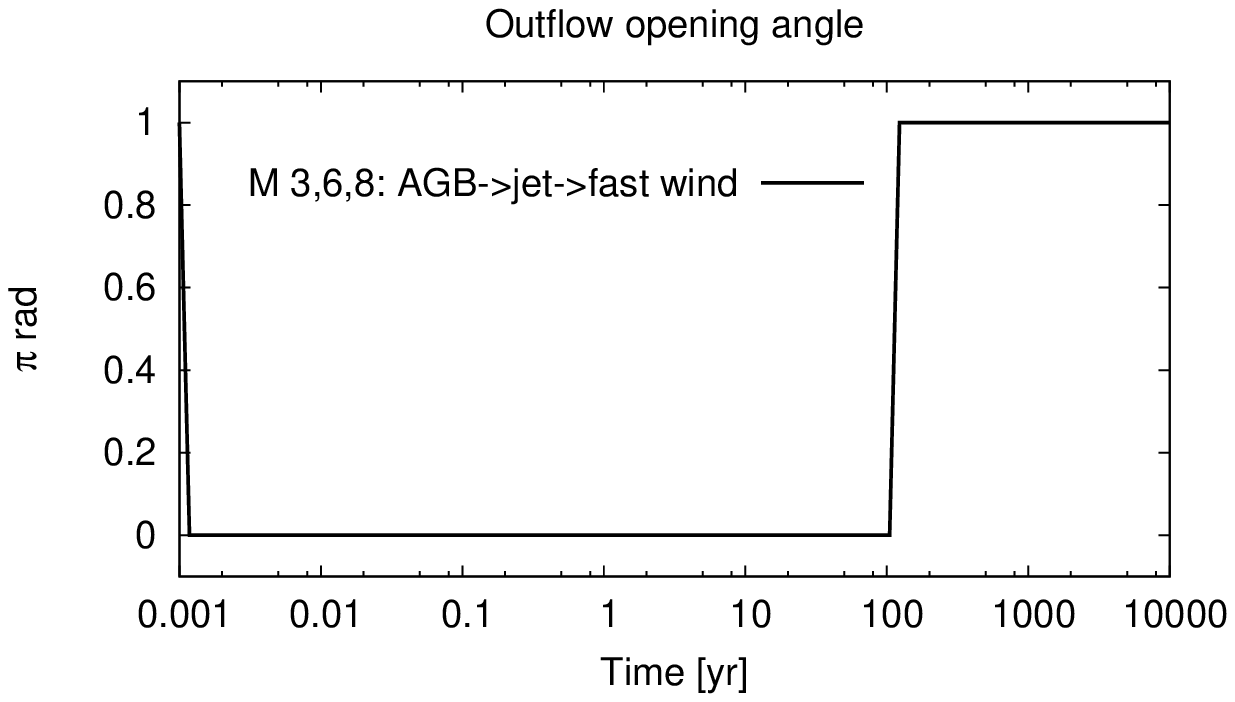}
\caption{
Outflow evolutionary profiles of Models~1-8.  The initial conditions (t$\le$0.001\,yr)
are those of the AGB wind and are set throughout the computational
domain, whereas the conditions of both jet and fast wind phases
are set only on the gas located inside the nozzle. See Table~1
for details.
}   \label{profiles}
\end{figure*}

All simulations and their parameters, 
% 
%%%JOEL 
   including the additional models that we present below, 
%%% 
% 
are summarized in Table~1.
Evolutionary profiles of the outflow conditions are presented in
Figure~\ref{profiles}. We note again that the initial conditions are
those of the AGB wind and set throughout the computational domain,
whereas the conditions of both jet and fast wind episodes are set
only on the gas located inside the nozzle ($R < r_w$, see above).

%%%%%%%%%%%%%%%%%%%%%%%%%%%%%%%%%%%%%%%%%%%%%%%%%%%%%%%%%%%%%%%%%%%%
\setcounter{table}{0}
\begin{table*}
\centering
    \begin{minipage}{108mm}
\caption{Simulations and parameters. \label{t1}}
   \begin{tabular}{@{}lccccc@{}}
\hline           
Simulation    &AGB wind    &Jet                 &Fast wind        &Fast wind               &Ionization             \\
name          &distribution   &duration            &duration    &max. speed         &front       \\
~~~~          &~~~~   &$[\times$100\,yr$]$   &$[\times$1000\,yr$]$    &$[\times$1000\,km\,s$^{-1}]$   &passage \\
\hline
\one        &spherical              &1                &~0.0   &\ldots   &n \\
\one*       &spherical              &1                &~0.0   &\ldots   &y \\
\two        &spherical              &0                &13.0   &2        &n \\
\two*       &spherical              &0                &13.0   &2        &y \\
\twoc       &spherical              &0                &16.3   &1        &n \\
\twoc*      &spherical              &0                &16.3   &1        &y \\
\three      &spherical              &1                &10.7   &2        &n \\
\three*     &spherical              &1                &10.7   &2        &y \\
\four       &toroidal~$\rho$        &1                &~0.0   &\ldots   &n \\
\five       &toroidal~$\rho$        &0                &~3.8   &2        &n \\
\six        &toroidal~$\rho$        &1                &~6.0\footnote{
This time is shorter than the one of \three\
(10.7\,kyr) because the toroidal AGB wind in this model funnels the nebula's lobe 
and thus it reaches the computational domain's boundary faster than the lobe 
in \three.}
&2        &n \\
\seven      &aspherical~$\vec{v}$   &0                &13.0   &2        &n \\
\eight\footnote{\eight\ includes a relaxation interlude of
$\sim\,$400\,yr between the jet and the wind episodes, and a computational
domain of 2\,pc$^2$, instead of 1\,pc$^2$ as in other simulations.}
&spherical      &1                &18.0   &2        &n \\
   \hline
   \end{tabular}
   \end{minipage}
\label{table1}
\end{table*}
%%%%%%%%%%%%%%%%%%%%%%%%%%%%%%%%%%%%%%%%%%%%%%%%%%%%%%%%%%%%%%%%%%%%

%%%%%%%%%%%%%%%%%%%%%%%%%%%%%%%%%%%%%%%%%%%%%%%%%%%
%%%%%%%%%%%%%%%%%%%%%%%%%%%%%%%%%%%%%%%%%%%%%%%%%

%%%%%%%%%%%%%%%%%%%%%%%%%%%%%%%%%%%%%%%%%%%%%%%%%%%%%
\subsection{Ionization front} \label{ionfront}

As the central star evolves from the AGB to the proto-white dwarf stages its surface
temperature 
   dramatically and rapidly %joel 
increases and so does its flux of ionizing UV photons 
\citep{balick87,hollenbach97,
perinotto04}. %Balick
Thus an ionization front will be driven away from the star into the
surrounding circumstellar medium.  Upon ionization the former neutral AGB 
envelope will be heated
from temperatures of order $T_{neb} \sim 10\,K$ to $T_{neb} \sim 10^4\,K$.

The numerical description of ionization fronts with AMR requires
the implementation of sophisticated methods for radiative transfer.
Efforts are currently underway to incorporate such methods into AstroBEAR2.0.
In the current study we have implemented a simplified
approximation which accounts for the  thermal effect of ionization.
When the fast stellar wind reaches a velocity of 2000\,km\,s$^{-1}$
we increase the temperature of cold gas~($\la$500\,K) to~10000\,K. This temperature
increase is imposed instantaneously  across the entire computational
domain.  In this way we assume that the nebula is density bounded
and that the ionization front is R-type\footnote{
% 
%%% SAHAI 
   In a R-type ionization front gas is advected at supersonic velocities 
   and exhausts at a slightly lower, but still
   supersonic, velocity.} \citep[see][and references therein]{PNfronts},
%%% 	
sweeping across the nebular
domain in a timescale $t_i \sim R/c$ that is very short compared with the
hydrodynamical evolution timescale $t_h \sim R/v_w$ (where $R, c$
and $v_w$ are the radius of the nebula, the speed of light and the
speed of the fast wind, respectively).
% 
%%%JOEL 
   In reality, however, ionization will not be as effective in the
	equatorial plane as towards the poles.
%%% 

Simulations that 
% 
%%%JOEL 
include this parametrization of ionization front passage %use this method 
%%% 
% 
are marked with an asterisk.  For
example, other than including the 
ionization-generated %JOEL 
temperature increase, \one*
is identical to \one, \two* is identical to \two, etc. (see
Table~\ref{t1}). While this method is crude it allows us
to test 
%ORSPLA may2011 
   %the only facet of ionization that matters for the present study, i.e. 
% 
%%%JOEL 
   %which focuses on global changes in morphology: 
	the potential
%%% 
of the pressure generated  via ionization to change the evolution
of the overall nebular morphology at late states in its evolution.

%%%%%%%%%%%%%%%%%%%%%%%%%%%%%%%%%%%%%%%%%%%%%%%%%
\subsection{Additional models} \label{aspherical}

For completeness we have also chosen to carry out a series of GISW
models in which bipolar nebular morphologies are expected to form
in the PN phase (rather than through jets in the
PPN phase).  The GISW model relies on pole to equator density
contrasts in an aspherical, toroidally shaped AGB envelope.  Since
toroidal AGB density distributions have been adopted in GISW models
before (e.g. see \citealp{frank94}) %huggins07}) ORSOLA
we have decided to run
a series of models that include such density gradients as initial
AGB wind conditions. Specifically, the AGB (ambient/circumstellar) 
density distribution in Models~4, 5~and~6 
is the AGB aspherical density distribution 
used by \citet[][but see also references therein]{icke92} 
and \citet{frank94}, where
%(AF: Cite correctly ICKE USED THIS FIRST, in ICKE, BALICK FRANK or before.  Then cite Frank \& Mellema)

\begin{eqnarray}
   \rho_{torus} & = & \frac{\rho_{AGB}}{F(\theta)}
      \left( \frac{r_w}{r} \right)^2, \\
   F(\theta) & = & 1 - \alpha \left(
      \frac{e^{\beta cos(2 \theta) - \beta -1}}{e^{- 2 \beta}-1} \right),
\label{torusEQ}
\end{eqnarray}
\noindent and $\rho_{AGB} = \dot{M}_{AGB}/$ (4$ \pi
v_{AGB} r^2$).  

This functional form produces a well characterized pole to equator
density contrast.  For example with $\alpha=\,$0.5 and $\beta=\,$1
the above expressions introduce a pole to equator density contrast
of 0.5 in the AGB wind.  We have chosen these values in our study
as they have been shown to produce butterfly type bipolar morphologies
in pure GISW models (Frank \& Mellema 1994).

\four\ is a two phase simulation and follows the interaction only of
the jet with the toroidal AGB wind. \five\ is also a two phase
simulation but it follows only the fast wind interacting with the
toroidal AGB wind. \six\ is a three phase simulation tracking the
jet followed by the fast wind interacting with the (preexisting) toroidal AGB
wind (see Table~\ref{t1}).

Finally, %for completeness %%BALICK
we explore the role of velocity gradients
in the AGB wind.  We have carried out a simulation in~\seven\ in
which $V_{AGB} = V(\theta)=f(r,z)*V_o$ where the constant radial velocity field $V_o$ is multiplied by  the  function

\begin{equation}
   f(r,z) = 1+e^{-tan^{-1}(|r/z|)^2/0.3^2},
   \label{velField}
\end{equation}
%%%SAHAI 
   %which introduces a continuous pole to equator velocity contrast 
   %of~2 
\noindent which results in a latitudinal decreasing expansion velocity,
   with a polar-to-equatorial contrast ratio of 2.
%%% 
A continuous fast wind is then driven into this aspherical
AGB wind.  As we shall see in the next section such a velocity
gradient does not greatly affect the morphology
of the resulting nebula.

%%%%%%%%%%%%%%%%%%%%%%%%%%%%%%%%%
\section{RESULTS} \label{results}

The results of our simulations are presented via logarithmic false color gray scale maps 
of the gas density
with overlaid velocity vectors in Figures~\ref{a}--\ref{others}.
Panels in these figures are arranged such that 
columns correspond to different simulations and rows show 
time increasing from top to bottom.

%balick
   \subsection{Structure within spherical AGB winds} \label{3.1}

\begin{figure*}
 \centering
 \includegraphics[width=\textwidth]{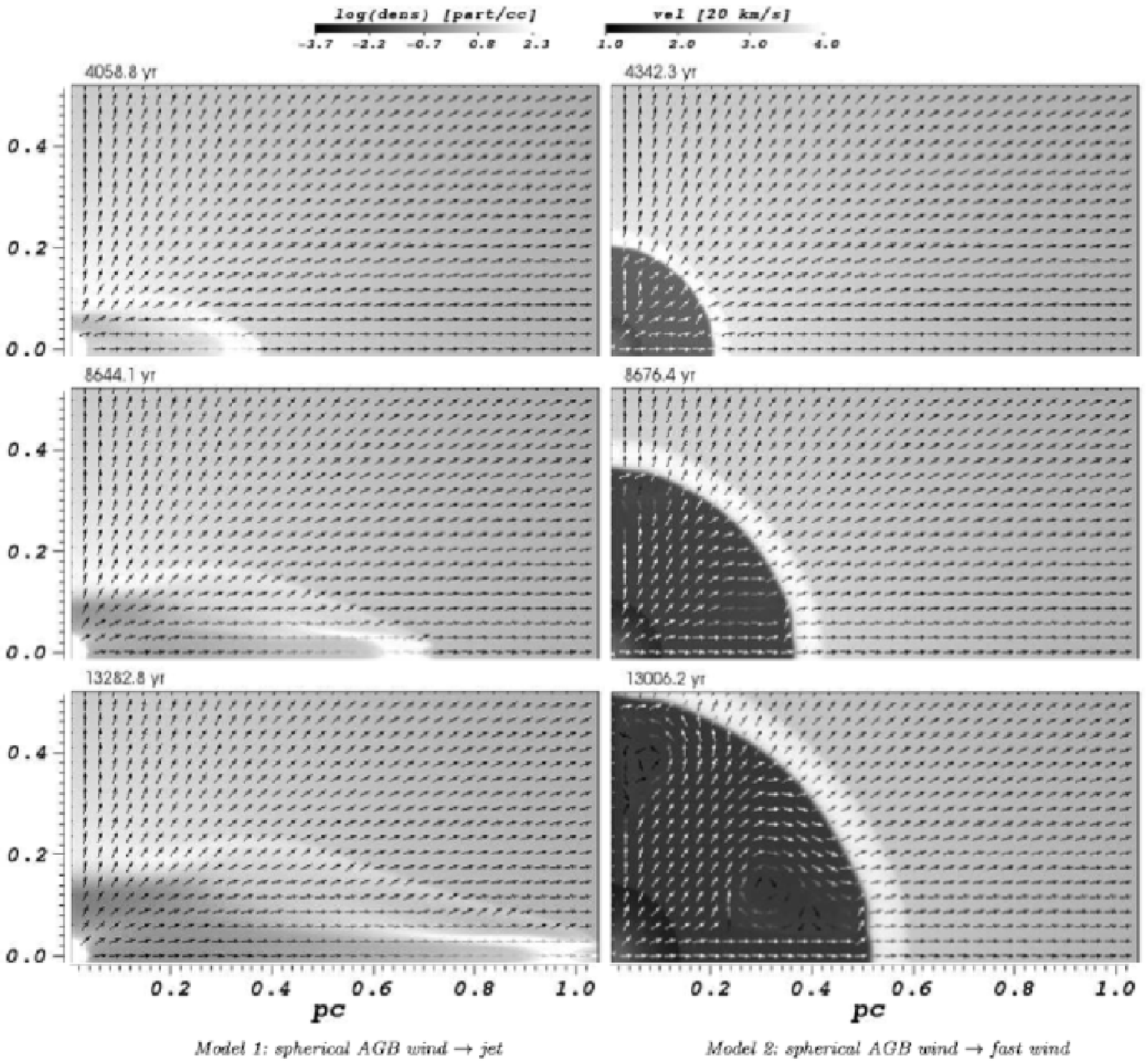}\\
%\includegraphics[width=.5\textwidth,bb=0.2in 8.8in  7.35in 9.6in,clip=]{scale.ps}\\
%%
%   \includegraphics[width=.065\textwidth,bb=1.00in 5.62in 1.86in 9.04in,clip=]{211.ps} 
%   \includegraphics[width=.462\textwidth,bb=1.86in 5.62in 7.9in  9.03in,clip=]{211.ps} 
%   \includegraphics[width=.462\textwidth,bb=1.86in 5.62in 7.9in  9.03in,clip=]{221.ps} \\
%	\hskip0cm $\downarrow$ \hskip9cm $\downarrow$ \\
%   \includegraphics[width=.065\textwidth,bb=1.00in 5.62in 1.86in 9.04in,clip=]{211.ps} 
%   \includegraphics[width=.462\textwidth,bb=1.86in 5.62in 7.9in  9.03in,clip=]{212.ps}
%   \includegraphics[width=.462\textwidth,bb=1.86in 5.62in 7.9in  9.03in,clip=]{222.ps} \\
%   \includegraphics[width=.065\textwidth,bb=1.00in 5.62in 1.86in 9.04in,clip=]{211.ps} 
%   \includegraphics[width=.462\textwidth,bb=1.86in 5.62in 7.9in  9.03in,clip=]{213.ps} 
%   \includegraphics[width=.462\textwidth,bb=1.85in 5.62in 7.9in  9.03in,clip=]{223.ps} \\
%~~~~~~~~~~~\includegraphics[width=.462\textwidth,bb=1.85in 1.95in 8.0in 2.70in,clip=]{223.ps} 
%          \includegraphics[width=.462\textwidth,bb=1.85in 1.95in 8.0in 2.70in,clip=]{223.ps} \\
%    \textit{\hskip2cm \one: spherical AGB wind $\rightarrow$ jet \hskip3cm
%    %$\qquad \qquad \qquad$
%    \two: spherical AGB wind $\rightarrow$ fast wind}
\caption{Dynamical evolution of the gas for different simulations
 that started with a spherical AGB wind distribution. \captA}
    \vspace*{0pt}
\label{a}
\end{figure*}

  %{\bf Jet Only Models}:
  {\bf Jet Models}:  In figure \ref{a}, left column, we show the
  results 
     of \one, %for a simulation 
  in which a short collimated jet phase
  is driven into a spherical AGB wind envelope.  The figure shows
  the jet driving a bow~shock which bounds a narrow-waisted cavity.
  The cavity wall is defined by a thin dense ``rim'' and its bipolar
  geometry persists for the entire duration of the simulation. The
  lateral width of the lobe is the result of the thermal pressure
  within the cavity as shocked material is forced laterally out of
  the region of the jet head. Thus the lobe's width-to-length ratio
  is determined by the sound speed in the AGB shell and the speed
  of the jet head.  We note that the cooling time of the post-shock
  gas, $t_c \sim T/(n \Lambda(T))$, 
     where $\Lambda(T)$ is the cooling function \citep{dm}, %ORSOLA
  is short relative to the hydrodynamic time
     (which is the ratio of the computational domain length %ORSOLA
	  over the gas sound speed). %ORSOLA
  Thus all post-shock flows in this simulation are relatively isothermal.

% 
%%% SAHAI AND MARTIN 
   After about 3500\,yr, the cavity lobe develops a narrow waist.
	This timescale is of order the expansion time of mature,
	but not old, PN.
   The pinch at the waist is the result of the higher density (and,
   therefore, higher ram pressure) of the AGB shell at small radii.
	In section~\ref{ratio} we analyze the evolution of the
	aspect ratio of some of our model nebulae. Here we note the
	aspect ratio of the cavity's bounding bipolar lobes increase monotonically in
	time.
%%% 

 %{\bf Fast Wind Only Models}:
 {\bf Fast Wind Models}: In figure~\ref{a}, right column,
 we show the results of \two,
 in which a ``classical'' fast wind is
 driven into a spherical AGB wind envelope. The growth of the nebula
 follows the  analytical predictions of Kwok et~al. (1978, 
 \citealp[but see also][]{stute06}) and %SAHAI 
 %balick, feb2011 
    agrees well with Sch\"onberner, Jacob, \& Steffen (2005),
	 whose 1-D models include much more detailed physics that
	 we could include in our multidimensional models.
 %%%
 The slow dense AGB wind
 is quickly overtaken by the spherical fast wind which is initially
 100~times 
    sparser %lighter %%BALICK
 and 20~times faster. An inward facing shock
 decelerates fast wind gas which is then heated to temperatures of
 $T \sim 10^7-10^8$K; a spherical hot bubble is formed.
 % 
 %%%JOEL 
    Such high temperatures have never been observed in PN hot bubbles,
	 and, hence, a number of mechanisms have been proposed to moderate their
	 temperatures (\citealp[e.g. see][and references therein]{soker10}; \citealp{shuleli2012}).
 %%% 
 The thermal pressure of the hot bubble acts as the piston driving a shock into the AGB
 envelope.  AGB gas swept~up by this shock cools rapidly and is
 compressed into a dense rim expanding with a velocity of $V_{rim} \sim\,$
 20\,km\,s$^{-1}$. In this way an isotropically expanding PN is formed,
 % 
 %%%JOEL 
    just as described by the simple analytical model of \citet{kwok78}.
 %%% 	 
 The nebula reaches a radius of $\sim\,$0.4\,pc after expanding for
 about~5000\,yr.

We note that observations (Bruce Balick 2011, private communication)
%Balick feb2011
	% 
	%%% Sch\"onberner, email, 10sept 2011 quote:
	   %"... misinterpretation: we are talking about ionised nebular shells
	   %in which photo-ionisation established positive velocity
	   %gradients (roughly v ~ r)."
		%
         %and also the hydrodynamical models of Sch\"onberner, Jacob, \& 
	      %Steffen (2005)
   %%% 
	% 
%%%
often find that AGB shells expand with an
outward radial gradient  (i.e. $v_{AGB} \propto R_{AGB}$). This is
not the case in our AGB initial conditions (section~\ref{epis}).
However, we carried out additional test simulations (not shown) and
have found that 
%blakman, 12 feb 2011
  %the shape of our PN rims are not affected by AGB
  %environments in which $v_{AGB} \propto R_{AGB}$.
  the morphology results of the present paper do not sensitively
  depend on the functional form, $v_{AGB}=~$constant vs $v_{AGB}
  \propto R_{AGB}$, of AGB environments. 
%%% 
%
%%%MARTIN, BASE ON ORSOLA'S COMMENTS:
   This happens, in part, because the velocity of the jet and fast wind is
   faster than $v_{AGB}$ by at least one order of magnitude.
%%%

\begin{figure*}
 \centering
 \includegraphics[width=\textwidth]{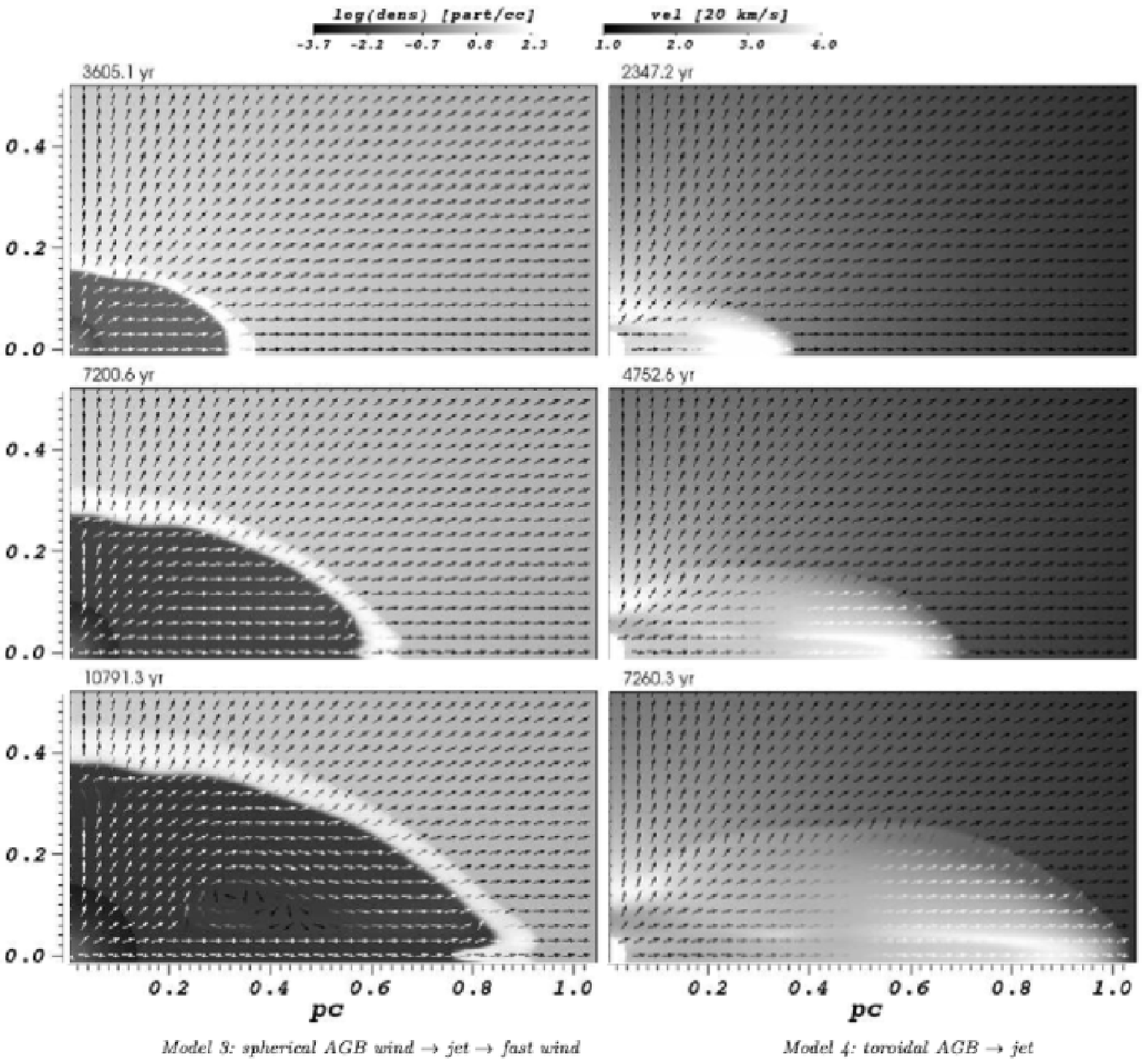}\\
%\includegraphics[width=.5\textwidth,bb=0.2in 8.8in  7.35in 9.6in,clip=]{scale.ps}\\
%%
%   \includegraphics[width=.065\textwidth,bb=1.00in 5.62in 1.86in 9.04in,clip=]{211.ps} 
%   \includegraphics[width=.462\textwidth,bb=1.86in 5.62in 7.9in  9.03in,clip=] {231.ps} 
%   \includegraphics[width=.462\textwidth,bb=1.86in 5.62in 7.9in  9.03in,clip=] {311.ps} \\
%   \includegraphics[width=.065\textwidth,bb=1.00in 5.62in 1.86in 9.04in,clip=]{211.ps} 
%   \includegraphics[width=.462\textwidth,bb=1.86in 5.62in 7.9in  9.03in,clip=] {232.ps}
%   \includegraphics[width=.462\textwidth,bb=1.86in 5.62in 7.9in  9.03in,clip=] {312.ps} \\
%   \includegraphics[width=.065\textwidth,bb=1.00in 5.62in 1.86in 9.04in,clip=]{211.ps} 
%   \includegraphics[width=.462\textwidth,bb=1.86in 5.62in 7.9in  9.03in,clip=] {233.ps} 
%   \includegraphics[width=.462\textwidth,bb=1.85in 5.62in 7.9in  9.03in,clip=] {313.ps} \\
%~~~~~~~~~~~\includegraphics[width=.462\textwidth,bb=1.85in 1.95in 8.0in 2.70in,clip=]{223.ps} 
%          \includegraphics[width=.462\textwidth,bb=1.85in 1.95in 8.0in 2.70in,clip=]{223.ps} \\
%    \textit{\hskip1cm 
%   \three: spherical AGB wind $\rightarrow$ jet $\rightarrow$ fast wind
%    \hskip3cm
% \four: toroidal AGB $\rightarrow$ jet}
 \caption{Dynamical evolution of the gas.  \captA}
    \vspace*{0pt}
\label{a2}
\end{figure*}

{\bf Jet and Fast Wind Models}:  In figure \ref{a2}, left column, we
show the results of \three,
in which the collimated jet phase is
followed by the spherical fast wind.  Initially, the collimated jet
forms a bipolar cocoon at the center of the AGB envelope as in \one.
Once formed, the jet-driven cavity is then shaped from within
by the isotropic fast wind. The fast-wind shock quickly forms, filling
the initially bipolar cavity with high temperature gas. The contact
surface of the jet-driven cavity is pushed against the slow denser AGB
envelope by the internal pressure of the fast wind. Because the jet
has already shaped the AGB wind, an elliptical, rather than a
spherical, dense rim is eventually formed.

The bubble attains an aspect ratio of $\approx 2$ which is achieved early
in the evolution and remains roughly constant as the nebula
expands. This simulation demonstrates one of the principal points
of our study: \emph{An initially bipolar outflow driven by a strongly
collimated jet 
can %will %SAHAI 
be transformed into an elliptical nebula at
longer times and larger scales by a subsequent isotropic, fast wind}.  
Thus nebular morphology is not a given of initial conditions but can 
change over time as the wind driving conditions change.

In most of our models the transition between jet and wind occurs
instantly. In \eight\ (figure~\ref{others}, right column) however, 
we show the
results of a simulation in which we have modified the conditions
in \three\ to include a quiescent period between the jet and the wind.  In
the evolution of real PN this gap could represent a phase during
which energy injection from the CSPN, or binary system, may be
absent. As discussed in the previous sections time-scale estimates
of such episodes are of order the jet duty cycle, $t \sim\,$100\,yr
(see section~\ref{epis}). The simulation shows that, once again,
the jet phase leads to the formation of 
a cavity that is initially bipolar. %an initially bipolar cavity. %SAHAI 
This fossil structure then expands quasi-ballistically for~$\sim\,$400\,yr,
reaching scales of order~0.05\,pc.  The fast wind then begins at
time $t =\,$500\,yr, and the evolution of the interacting winds is
essentially similar to that of \three.

We note that in this simulation the rim of swept up AGB material maintains
an aspect ratio close to~2.3 which means that once the fast wind
begins the nebula expands homologously (see Figure~\ref{aspect}
and section~\ref{velo}).  One
difference between this simulation and \three\ is the late time
formation of a dense knot along the long axis of the elliptical rim.
The number density and flow speed  in the knot are of order~%$n =\,$
1500\,cm$^{-3}$ and~%$v =\,$
30\,km\,s$^{-1}$, respectively, which are not
unlike those in some FLIERS
% 
%%%JOEL
   \citep[see e.g.][and references therein]{balick98}.
%%% 
% 
The formation of the knots appears
due to the radiative collapse of jet material as it interacts with
the AGB wind. In contrast, when the fast wind immediately follows the jet phase
(\three) the flux of mechanical energy into the polar regions flattens the
jet material and keeps it from forming a ballistically expanding dense knot.
%%

%%%%%%%%%%%%%%%%%%%%%%%%%%%%%%%%%%%%%%%%%%%%%%
%%%%%%%%%%%%%%%%%%%%%%%%%%%%%%%%%%%%%%%%%%%%%%%
%%%%%%%%%%%%%%%%%%%%%%%%%%%%%%%%%%%%%%%%%%%%%%
%%%%%%%%%%%%%%%%%%%%%%%%%%%%%%%%%%%%%%%%%%%%%%%

%%%%%%%%%%%%%%%%%%%%%%%%%%%%%%%%%%%%%%%%%%%%%%
%\begin{figure*}
% \centering
% \includegraphics[width=.6\textwidth,bb=0.2in 8.8in  7.35in 9.6in,clip=]{scale.ps}\\
% %
%    \includegraphics[width=.353\textwidth,bb=1.00in 1.90in 8.0in 9.25in,clip=]{211.ps} 
%    \includegraphics[width=.31\textwidth,bb=1.85in 1.90in 8.0in 9.25in,clip=]{212.ps}
%    \includegraphics[width=.31\textwidth,bb=1.85in 1.90in 8.0in 9.25in,clip=]{213.ps} \\
%    \one: spherical AGB wind $\rightarrow$ jet \\
%    \includegraphics[width=.353\textwidth,bb=1.00in 1.90in 8.0in 9.25in,clip=]{221.ps}
%    \includegraphics[width=.31\textwidth,bb=1.85in 1.90in 8.0in 9.25in,clip=]{222.ps}
%    \includegraphics[width=.31\textwidth,bb=1.85in 1.90in 8.0in 9.25in,clip=]{223.ps} \\
%    \two: spherical AGB wind $\rightarrow$ fast wind \\
%    \includegraphics[width=.353\textwidth,bb=1.00in 1.90in 8.0in 9.25in,clip=]{231.ps}
%    \includegraphics[width=.31\textwidth,bb=1.85in 1.90in 8.0in 9.25in,clip=]{232.ps}
%    \includegraphics[width=.31\textwidth,bb=1.85in 1.90in 8.0in 9.25in,clip=]{233.ps} \\
%    \three: spherical AGB wind $\rightarrow$ jet $\rightarrow$ fast wind \\
% \caption{Dynamical evolution of the gas for different simulations
% that started with a spherical AGB wind distribution. \captA}
%    \vspace*{0pt}
%\label{a}
%\end{figure*}
%%%%%%%%%%%%%%%%%%%%%%%%%%%%%%%%%%%%%%%%%%%%%%%

%BALICK
  \subsection{Structure formation in non-spherical AGB shells} 
  \label{3.2}
\begin{figure*}
 \centering
 \includegraphics[width=\textwidth]{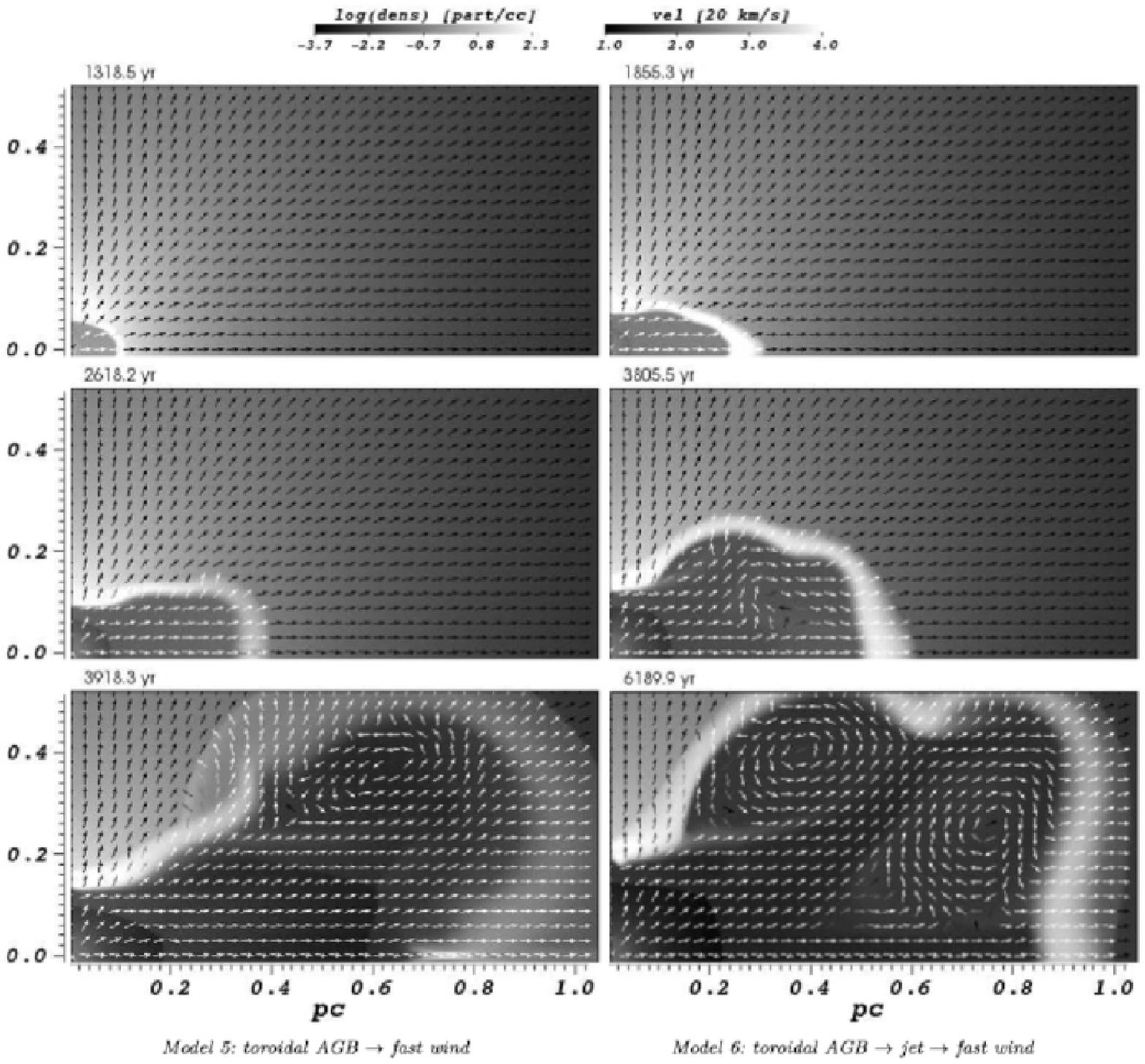}\\
%\includegraphics[width=.5\textwidth,bb=0.2in 8.8in  7.35in 9.6in,clip=]{scale.ps}\\
%%
%   \includegraphics[width=.065\textwidth,bb=1.00in 5.62in 1.86in 9.04in,clip=]{211.ps} 
%   \includegraphics[width=.462\textwidth,bb=1.86in 5.62in 7.9in  9.03in,clip=] {321.ps} 
%   \includegraphics[width=.462\textwidth,bb=1.86in 5.62in 7.9in  9.03in,clip=] {331.ps} \\
%   \includegraphics[width=.065\textwidth,bb=1.00in 5.62in 1.86in 9.04in,clip=]{211.ps} 
%   \includegraphics[width=.462\textwidth,bb=1.86in 5.62in 7.9in  9.03in,clip=] {322.ps}
%   \includegraphics[width=.462\textwidth,bb=1.86in 5.62in 7.9in  9.03in,clip=] {332.ps} \\
%   \includegraphics[width=.065\textwidth,bb=1.00in 5.62in 1.86in 9.04in,clip=]{211.ps} 
%   \includegraphics[width=.462\textwidth,bb=1.86in 5.62in 7.9in  9.03in,clip=] {323.ps} 
%   \includegraphics[width=.462\textwidth,bb=1.85in 5.62in 7.9in  9.03in,clip=] {333.ps} \\
%~~~~~~~~~~~\includegraphics[width=.462\textwidth,bb=1.85in 1.95in 8.0in 2.70in,clip=]{223.ps} 
%          \includegraphics[width=.462\textwidth,bb=1.85in 1.95in 8.0in 2.70in,clip=]{223.ps} \\
%    \textit{\hskip1cm 
%\five: toroidal AGB $\rightarrow$ fast wind 
%    \hskip3cm
%\six: toroidal AGB $\rightarrow$ jet $\rightarrow$ fast wind}
 \caption{Dynamical evolution of the gas for different simulations
   that started with a toroidal AGB wind density distribution.\captA}

    \vspace*{0pt}
\label{torus}
\end{figure*}

The nebular dynamics changes significantly when the AGB envelope
begins with a toroidal density distribution.  The results of our
studies in this context are summarized in Figure~\ref{torus}. We note that in these
models the jet propagates along the symmetry axis of the toroidal
AGB envelope.

  %{\bf Jet Only Model}:
  {\bf Jet Model}: In figure ~\ref{a2}, right column, we show
  the results of \four, in which the jet alone drives into an
  aspherical AGB wind.  Once again the jet drives a fossil bipolar
  cavity which develops a narrow waist at~$\sim\,$3000\,yr
  compared to 5000\,yr for the spherical AGB shell 
  (of \one).  The
  lateral expansion of the lobe is, once again, stronger in 
  %balick feb2011
     %regions beyond the densest part of the toroidal AGB envelope. 
	  %
     lower density regions where the ram pressure of the initial
	  AGB envelope is easier to displace. 
  %%%
  The lobe that develops in this case has a narrower waist than
  in models with no pole to equator density contrast. The
  dynamics of the evolution is, however, not significantly different
  %balick 
     %from what was seen in the model \one. 
	  from the results of \one.
  %%% 
  %orsola 
     This means 
	  % 
	  %%%joel 
	     %that PPN would look the same 
	     it would be difficult to determine from the morphology of a PPN
	  %%% 
	  whether or not the progenitor AGB wind had a significant
	  pole to equator density contrast. Jets, if present, dominate 
	  the morphology only if there is no fast wind or if the fast wind 
	  has less momentum.
  %%% 

 %{\bf Fast Wind Only Models}:
 {\bf Fast Wind Model}: In figure ~\ref{torus}, left column,
 we show the results of \five,
 in which only a fast light isotropic
 wind interacts with the dense slow toroidal AGB envelope.  This
 is the classic GISW model that has been explored before (e.g.
 Icke et al. 1992; Frank \& Mellema 1994). The
 Figure shows that, as expected, the
 swept-up AGB shell becomes elliptical at early times and then
 becomes more significantly bipolar as the hot bubble drives the
 shock to larger radii in the AGB wind along the polar axis.  At
 late times the dense rim takes on a butterfly morphology as would
 be expected for these initial conditions. Thus this simulation
 and that of \two\ (spherical AGB wind)
 recover the results of previous numerical experiments in the GISW 
 and ISW models \citep[e.g.][]{frank94} and 
    thereby confirm %joel
 the consistency of our numerical simulations.

%{\bf Jet and Wind Only Model}
{\bf Jet and Fast Wind Model}: In figure ~\ref{torus}, right column,
we show the results of \six, 
in which the jet is followed by the fast
isotropic wind as they interact with the  toroidal AGB
envelope. Once again an initially bipolar
cavity and rim are created by the jet.  Once the fast wind begins
however the evolution differs significantly from \three.
Rather than changing the morphology from bipolar to elliptical, the
action of the fast wind in this case is different; the evolution
is dominated not by reshaping the jet driven
cavity but by interacting with the toroidal AGB wind.  The long
term evolution of this model leads to a butterfly shaped rim which
is quite similar to that seen in the previous simulation in which
no jet was included. \textit{Thus it appears that the presence of a
preexisting density contrast in the AGB wind is more important
for the evolution of morphology than the presence of a
short-lived jet phase as no change from bipolar morphology to
elliptical morphology occurs.}
%balick, feb2011
   We note 
	a large fraction %about 50\% %JOEL 
	of optical PPN have dense dust waists
	\citep[see e.g.][and references therein]{huggins07}.
	%Bruce Balick's compilation of optical PPN %AGB ORSOLA
	%images\footnote{
	%http://www.astro.washington.edu/users/balick/pPNe/} and
	%references therein). 
	Hence the above conclusion would suggest
	that a large fraction of mature PN should also be bipolar. Since this is not the case it is possible that such dust shells may not be extended enough to influence PN morphology once the fast wind begins.
%%%

\begin{figure*}
 \centering
 \includegraphics[width=\textwidth]{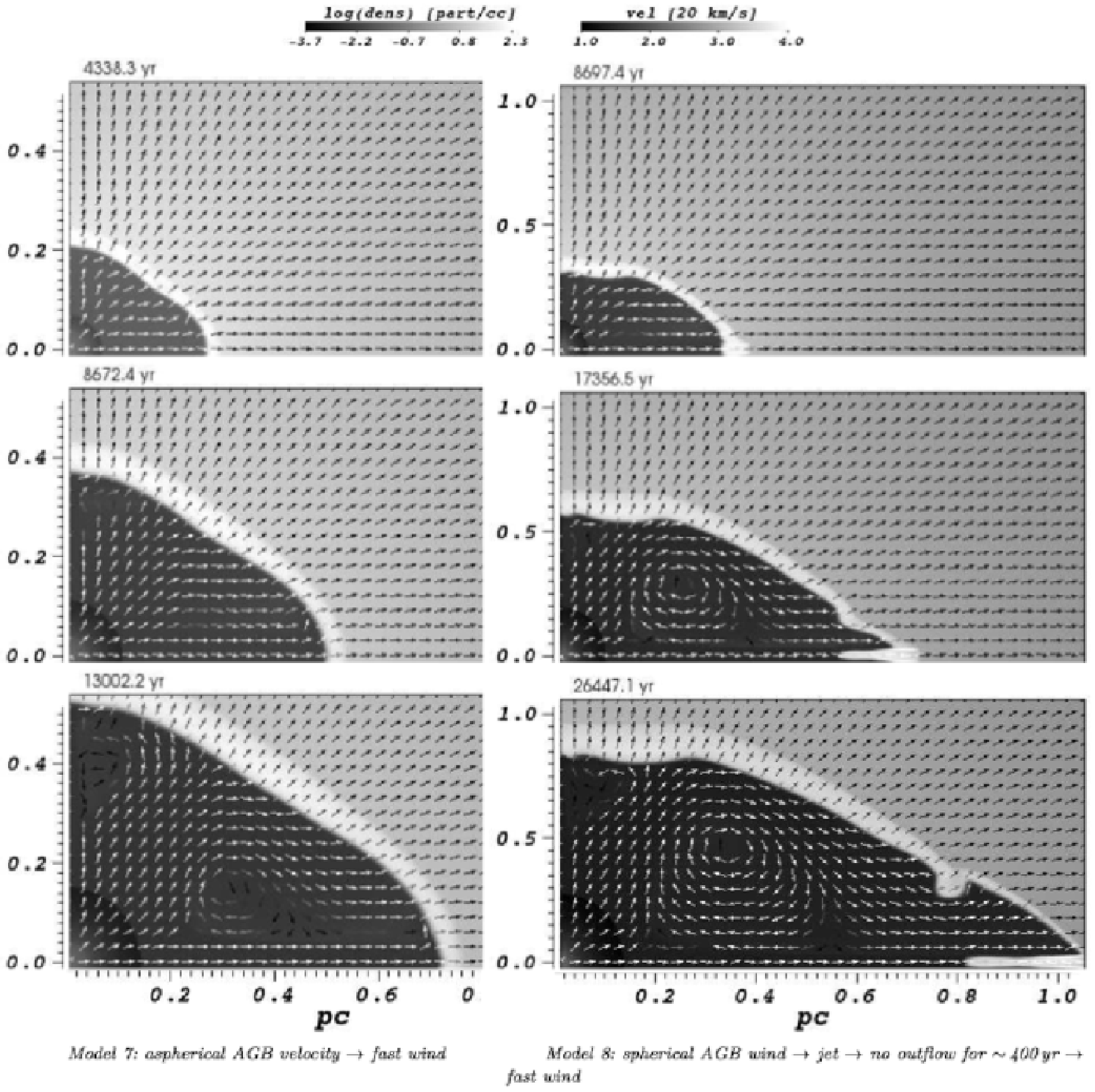}\\
%\includegraphics[width=.5\textwidth,bb=0.2in 8.8in  7.35in 9.6in,clip=]{scale.ps}\\
%%
%   \includegraphics[width=.059\textwidth,bb=1.10in 5.62in 1.86in 9.04in,clip=]{211.ps} 
%   \includegraphics[width=.377\textwidth,bb=1.86in 5.62in 6.5in  9.03in,clip=] {411.ps} 
%   \includegraphics[width=.059\textwidth,bb=1.10in 5.62in 1.86in 9.04in,clip=] {421.ps} 
%   \includegraphics[width=.483\textwidth,bb=1.86in 5.62in 7.9in  9.03in,clip=] {421.ps} \\
%   \includegraphics[width=.059\textwidth,bb=1.10in 5.62in 1.86in 9.04in,clip=]{211.ps} 
%   \includegraphics[width=.377\textwidth,bb=1.86in 5.62in 6.5in  9.03in,clip=] {412.ps}
%   \includegraphics[width=.059\textwidth,bb=1.10in 5.62in 1.86in 9.04in,clip=] {421.ps} 
%   \includegraphics[width=.483\textwidth,bb=1.86in 5.62in 7.9in  9.03in,clip=] {422.ps} \\
%   \includegraphics[width=.059\textwidth,bb=1.10in 5.62in 1.86in 9.04in,clip=]{211.ps} 
%   \includegraphics[width=.377\textwidth,bb=1.86in 5.62in 6.5in  9.03in,clip=] {413.ps} 
%   \includegraphics[width=.059\textwidth,bb=1.10in 5.62in 1.86in 9.04in,clip=] {421.ps} 
%   \includegraphics[width=.483\textwidth,bb=1.85in 5.62in 7.9in  9.03in,clip=] {423.ps} \\
%\hskip1.0cm \includegraphics[width=.377\textwidth,bb=1.85in 1.95in 6.5in 2.70in,clip=]{223.ps} 
%\hskip1.1cm\includegraphics[width=.483\textwidth,bb=1.85in 1.95in 8.0in 2.70in,clip=]{223.ps} \\
%    \textit{\hskip1cm 
%\seven: aspherical AGB velocity $\rightarrow$ fast wind
%    \hskip1.5cm
%\eight: spherical AGB wind $\rightarrow$ jet $\rightarrow$ no outflow for~$\sim\,$400\,yr
%$\rightarrow$ fast wind}
 \caption{Dynamical evolution of the gas.  \captA}

    \vspace*{0pt}
\label{others}
\end{figure*}

{\bf Asymmetric AGB Velocity:} In figure~\ref{others}, left column, we
present \seven, %a simulation 
in which the fast wind expands into an AGB
envelope with an aspherical velocity (rather than density) distribution.
In our initial conditions the AGB wind expands two~times faster
toward the pole than towards the equator, following equation (\ref{velField}).
The hot bubble responds to the reduced AGB ram pressure $P_{rp}(\theta)
= \rho_{AGB}V_{AGB}(\theta)^2$ along the equator leading to high
shock velocities at these latitudes.  
As the figure shows the resulting morphology is elliptical.
This simulation demonstrates that if 
% 
%%%ORSOLA 
   the early evolutionary phase of the PN produce a rapidly expanding
   {\it equatorial flow} rather than a {\it bipolar jet} \citep{nordhaus07},
   then an elliptical nebula might be the long term result.

%%%%%%%%%%%%%%%%%%%%%%%%%%%%%%%%%%%%%%%%%%%%%%%
%\begin{figure*}
% \centering
% \includegraphics[width=.6\textwidth,bb=0.2in 8.8in  7.35in 9.6in,clip=]{scale.ps}\\
% %
%    \includegraphics[width=.353\textwidth,bb=1.00in 1.90in 8.0in 9.25in,clip=]{311.ps} 
%    \includegraphics[width=.31\textwidth, bb=1.85in 1.90in 8.0in 9.25in,clip=]{312.ps}
%    \includegraphics[width=.31\textwidth, bb=1.85in 1.90in 8.0in 9.25in,clip=]{313.ps} \\
% \four: toroidal AGB $\rightarrow$ jet \\
% \vskip.2cm
%    \includegraphics[width=.353\textwidth,bb=1.00in 1.90in 8.0in 9.25in,clip=]{321.ps} 
%    \includegraphics[width=.31\textwidth, bb=1.85in 1.90in 8.0in 9.25in,clip=]{322.ps}
%    \includegraphics[width=.31\textwidth, bb=1.85in 1.90in 8.0in 9.25in,clip=]{323.ps} \\
% \five: toroidal AGB $\rightarrow$ fast wind \\
% \vskip.2cm
%    \includegraphics[width=.353\textwidth,bb=1.00in 1.90in 8.0in 9.25in,clip=]{331.ps} 
%    \includegraphics[width=.31\textwidth, bb=1.85in 1.90in 8.0in 9.25in,clip=]{332.ps}
%    \includegraphics[width=.31\textwidth, bb=1.85in 1.90in 8.0in 9.25in,clip=]{333.ps} \\
% \six: toroidal AGB $\rightarrow$ jet $\rightarrow$ fast wind
% \caption{Dynamical evolution of the gas for different simulations
%   that started with a toroidal AGB wind density distribution. \captA}
%    \vspace*{0pt}
%\label{torus}
%\end{figure*}
%%%%%%%%%%%%%%%%%%%%%%%%%%%%%%%%%%%%%%%%%%%%%%%
%%%%%%%%%%%%%%%%%%%%%%%%%%%%%%%%%%%%%%%%%%%%%%
%%%%%%%%%%%%%%%%%%%%%%%%%%%%%%%%%%%%%%%%%%%%%%%

%%%%%%%%%%%%%%%%%%%%%%%%%%%%%%%%%%%%%%%%%%%%%%%
%\begin{figure*}
% \centering
% \includegraphics[width=.6\textwidth,bb=0.2in 8.8in  7.35in 9.6in,clip=]{scale.ps}\\
% %
%    \includegraphics[width=.353\textwidth,bb=1.00in 1.90in 8.0in 9.25in,clip=]{411.ps} 
%    \includegraphics[width=.31\textwidth,bb=1.85in 1.90in 8.0in 9.25in,clip=]{412.ps}
%    \includegraphics[width=.31\textwidth,bb=1.85in 1.90in 8.0in 9.25in,clip=]{413.ps} \\
% \seven: aspherical AGB velocity $\rightarrow$ fast wind \\
%    \includegraphics[width=.353\textwidth,bb=1.00in 1.90in 8.0in 9.25in,clip=]{421.ps} 
%    \includegraphics[width=.31\textwidth,bb=1.85in 1.90in 8.0in 9.25in,clip=]{422.ps}
%    \includegraphics[width=.31\textwidth,bb=1.85in 1.90in 8.0in 9.25in,clip=]{423.ps} \\
% \eight: spherical AGB wind $\rightarrow$ jet $\rightarrow$ no outflow for~$\sim\,$400\,yr
% $\rightarrow$ fast wind
% \caption{Dynamical evolution of the gas. \captA}
%    \vspace*{0pt}
%\label{others}
%\end{figure*}
%%%%%%%%%%%%%%%%%%%%%%%%%%%%%%%%%%%%%%%%%%%%%%%
%%%%%%%%%%%%%%%%%%%%%%%%%%%%%%%%%%%%%%%%%%%%%%

\subsection{Effect of Ionization Induced Temperature Increase} \label{3.3}

We note finally that we have compared the evolution of those models with
our simple estimate of ionization against the ones with no change in
temperature (see Table~1, last column). We find only small changes,
of order 5\%, %martin 17feb2011
in the dynamical behavior
of simulations with a temperature increase (from $T \la\,$500\,K to
$T = 10^4$\,K).  
%
%balick: figure is not essential
  %We find the effect is confined to the width of the
  %swept up rim. Figure~\ref{ion} shows a cross cut of the rim in Models~2c
  %and~2d.  The width of the rim is only slightly larger in the
  %case with ionization as is to be expected due to higher thermal
  %pressures $P \propto T$ there.
  %%%%%%%%%%%%%%%%%%%%%%%%%%%%%%%%%%%%%%%%%%%%%%%
  %\begin{figure}[ht]
    %\centering
    %   \includegraphics[width=.50\textwidth]{ionVSno-ion-dens.ps}
    %\caption{Large-scale effects of our crude estimate of ionization
    %front passage on the fast spherical wind with a maximum wind
    %velocity of 1000\,km\,s$^{-1}$ (see Section~\ref{ionfront}). The
    %curves are radial profiles of the gas density, at 
    %$t =\,$16269\,yr, in \twoc\ (black curve) and \twoc* (red curve).
  %}
    %   \vspace*{0pt}
  %\label{ion}
  %\end{figure}
  %%%%%%%%%%%%%%%%%%%%%%%%%%%%%%%%%%%%%%%%%%%%%%%%

Our models are relevant to cases in which the ionization front remains
R-type (see section~\ref{ionfront}) until it leaves the AGB material.   
We are not able to see
details of the ionization front's passage on the gas in terms of
driving instabilities which can fragment the dense rim when the
fronts stall in the nebular gas.  We note that even 1-D models show
that the trapping of the ionization can have an effect on the
evolution of the rim  (\citealp{schomberner05}).  These effects
should not however affect our conclusions about global changes in
morphology.  For AGB envelopes that begin with little pole to equator density
contrast the trapping of the D-front will occur at all latitudes
and its changes will affect the entire rim. 
%balick
	%%Our simulations focus on global changes in the driving that occurs inside the rim.  
Thus ionization effects should not change our conclusions about the
morphological shift from bipolar to elliptical nebulae as they will
not change the nature of hot bubble pressure gradients driving the
rim from the inside.
We leave a detailed study on this matter for future work.

%%%%%%%%%%%%%%%%%%%%%%%%%%%%%%%%%%%%%%%%%%%%%%%
\section{DISCUSSION} \label{discu}

The main results from our simulations are the following. A young
bipolar PPN transforms into an older elliptical PN when an initial
spherical AGB envelope interacts with a short duration jet
and then with a spherical post-AGB fast wind.  Once mature PN
become elliptical they do not eventually change into
more spherical nebulae over relevant time scales. If the initial 
AGB envelope is toroidal then butterfly nebula occur with or without a PPN jet phase.
\subsection{Aspect ratio evolution} \label{ratio}

\begin{figure}
  \centering
     \includegraphics[width=.50\textwidth]{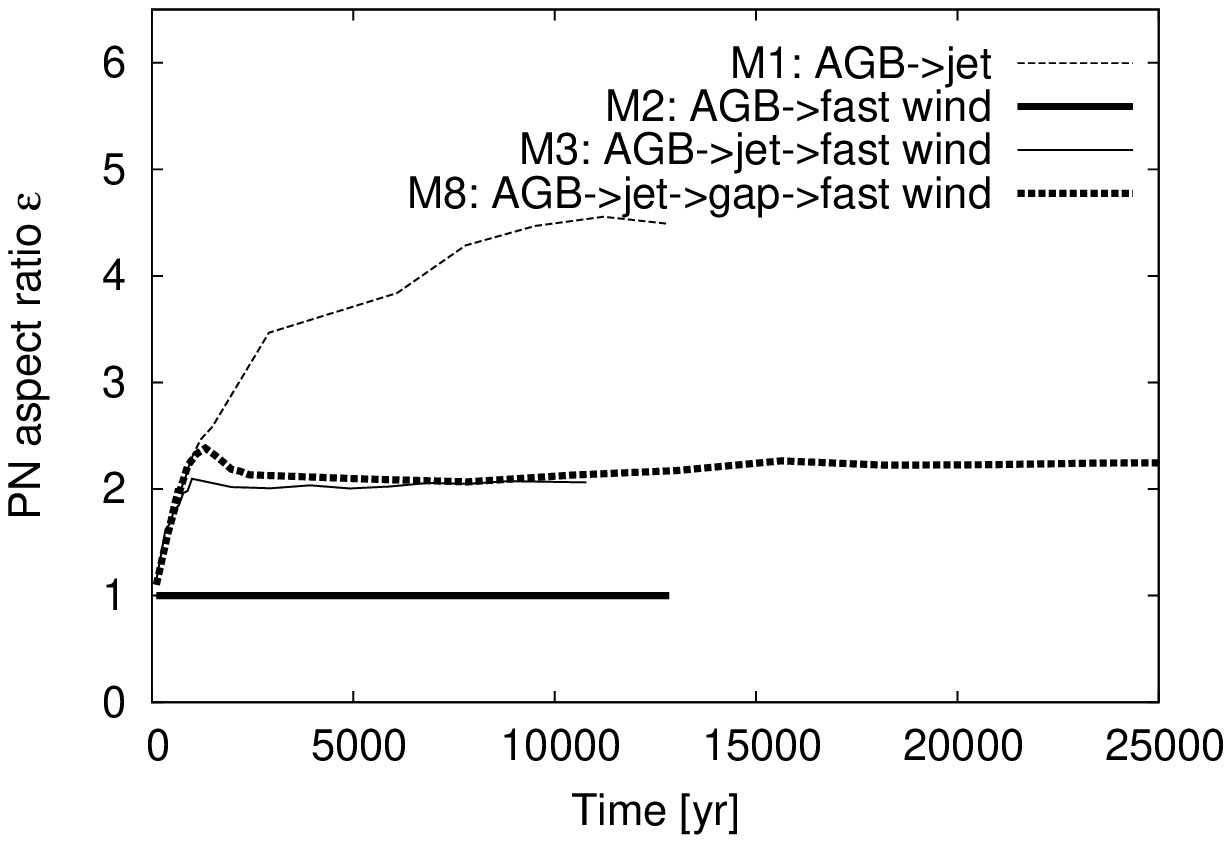}
  \caption{Time evolution of the nebular aspect ratio, $\epsilon(t)$.
The $y-$axis shows the major-to-minor axis ratio of the 
rims (the principal bright structures). 
Profile gradients correlate with the outflow phase sequence, or history.
}
     \vspace*{0pt}
\label{aspect}
\end{figure}

We follow the morphological transition of our model 
nebulae formed from spherical AGB envelopes 
(Figure~\ref{a}, Figure~\ref{a2}-left and Figure~\ref{others}-right)
using plots of the rims' aspect ratio,
$\epsilon(t) = L_{ma}(t)/L_{mi}(t)$, as a function of time. 
This is the major-to-minor
axis ratio of a rectangle that would contain the principal bright
structure of the nebulae.
%longer to shorter axis. $L_l(t)$ and $L_s(t)$ are measured at $r=\,$0 and
%$z=\,$0, respectively, and 
%We track $L_{ma}(t)$ and $L_{mi}(t)$ as a function of
%blackman, feb2011
   %the object's growth, $L_l(t)/L_l(t_{end})$, where $L_l(t_{end})$
   %is the corresponding rim's final longer axis. 
%	time.
%%%
%We do not compute $\epsilon(t)$ for narrow-waist nebulae
%(Figure~\ref{torus}) because it is not well defined.
The aspect ratio evolution profiles are shown in Figure~\ref{aspect}.  It is
clear that their gradients are related to the outflow phase
sequence, or history, in the simulations.

The spherical PN rim that we see in \two\ has a constant
$\epsilon(t)$ of unity (thick solid line). The bipolar
rim in \one\ has an $\epsilon(t)$ that increases monotonically
as the object expands (thin dashed line, Figure~\ref{aspect}).  
The expansion of
this lobe is dominated by the momentum of the jet during the early
evolutionary phases, hence we see a steep increase in $\epsilon(t)$. 
%balick
  %Then the steepness of $\epsilon(t)$ decreases as the expansion of the lobe becomes thermal, i.e. isotropic.
%adam 7 dec 2010
   Note that the slope decreases in time because the jet turns off.
   If the jet were active we would see a constant slope.
   The aspect ratio of the lobes is (of course) also proportional to 
   the duration of the jet injection phase.

In contrast, the rims in Models~3 and~8 (the thin solid and thick dashed lines
in Figure~\ref{aspect}, respectively)
show quite different evolution of $\epsilon(t)$.  
For these simulation $\epsilon(t)$ increases 
steeply during the early evolutionary
phases in which the jet dominates the object's expansion. This
corresponds to the PPN phase. Any further increase of the aspect ratio is then
quickly suppressed by the effects of the spherical fast wind. The
profiles reach values of about~2 and vary only modestly
afterwards.  This means the expansion of the rim becomes homologous
(section~\ref{velo}). This is the most important point of the 
work.

%%%%%%%%%%%%%%%%%%%%%%%%%%%%%%%%%%%%%%%%%%%%%%%
%%%%%%%%%%%%%%%%%%%%%%%%%%%%%%%%%%%%%%%%%%%%%%%

The thick dashed line corresponds to \eight\ and shows a short
increase/decrease in aspect ratio that occurs between the early
evolution and the later phases when the aspect ratio becomes constant.
This feature occurs during the quiescent evolutionary phase, i.e.
the PPN-PN transition.  Comparing with the aspect ratio profiles
of Models~1 and~3 we see that the height
of the bump in \eight\ is proportional to the duration of the
quiescent interlude  
%. Our choice of~$\sim\,$400\,yr is arbitrary; 
%about~4 times longer than the jet's duty cycle 
(see end of section~\ref{epis}).
An interesting question here is that of the dependence of $\epsilon(t)$
on the duration of this interlude. To answer this, additional
simulations (designed to explore a set of values for the duration
of the interlude) would have to be carried out; we leave
this for future research.

Our calculations therefore demonstrate how the morphology of PPN
and PN is correlated with the collimation parameters of the wind 
from the central object. The presence of a brief, hypersonic and heavy jet prior to the
interaction between the AGB envelope and the spherical fast wind
changes the shape of the resulting nebulae. The effects are important
during both the early and late phases of nebular expansion.  Bipolar
PPN are formed and then turn into elliptical PN, in 
broad agreement with the 
general statistical trends 
of the observed morphologies of PN.

%%%%%%%%%%%%%%%%%%%%%%%%%%%%%%%%%%%%%%%%%%%%%%%%%%%%
\subsection{Velocity field} 
\label{velo}

We show the velocity field of gas in our simulations
in Figures~2, 3, 4 and~5.
Vector density, location and length are constant, but vector
inclination and grey scale are related to the local direction and
speed of the flow, respectively. Arrow properties were chosen in
such a way as to stress velocity differences between the polar and
equatorial parts of our model nebular rims. A detailed study of the
velocity field distribution of material inside the lobes, as well
as in butterfly-shaped rims, is beyond the scope of this paper.

The following findings are consistent with the nebular aspect ratio
evolution profiles of Figure~\ref{aspect}.  The left column of
Figure~\ref{a} shows that the tip of the bipolar rim formed in Model~1
maintains a longitudinal (horizontal) velocity, $v_z$, which is close to
75\,km\,s$^{-1}$. Lobe material located behind the tip shows
a $v_z$ increase proportional to the distance from the star, $z$. This
is consistent with the ``Hubble flow'' kinematics observed in some
PPN outflows (Balick \& Frank 2002; Bujarrabal et al. 1998;
Olofsson \& Nyman 1999). In contrast, the velocity of the rim's waist
(Figure~\ref{a}, left)
is mostly transverse, $v_r$, slower (darker) and quite similar to that of the
AGB wind material located at larger radii. We see a similar velocity
distribution at the tip and waist of the bipolar rim formed in Model~4
(Figure~\ref{a2}, right column). I.e. a fast (light-gray) tip
with $v_r,v_z\sim\,$0,100\,km\,s$^{-1}$, and a slower
(dark-gray) waist with $v_r,v_z\sim\,$20,0\,km\,s$^{-1}$.  In
this case however most of the lobe material, not just the component
behind the rim's tip, also shows a $v_z$ increase
proportional to the distance from the star.

The left column of Figure~\ref{a2} shows that the elliptical PN rim
of Model~3 has a consistent, progressive speed increase (dark to
light gray) of a factor $\sim\,$2, form the waist to the tip along
the rim. This velocity field goes from transversely dominated at
small $z$ to longitudinally dominated at large $z$. The object
expands homologously.  In this case we do not see any correlation
between the position on the $z$-axis and the speed of the lobe material. 
We see similar
velocity distributions in the elliptical PN rims of Models~7 and~8
(Figure~\ref{others}). The tip-to-waist speed ratio $[v_z(r=0)/v_r(z=0)]$
of the rims in Models~7 and~8 is $\sim$\,1.5 and~2, respectively.
Also, we consistently find velocity vortices inside the lobes of our
model elliptical PN.
It is worth noting the semi-analytical kinematic analysis of 
\citet{steffenAPN5} which suggests that in general PN rims 
asymptotically evolve towards a homologous expansion, in agreement 
with our findings.

%%%%%%%%%%%%%%%%%%%%%%%%%%%%%%%%%%%%%%%%%%%%%%%%%%%%
\subsection{Synthetic observations} 
\label{obs}

%%% JOEL
   The simulations described in section~3 yield detailed predictions
   of the density structures of PN (as well as their detailed, if
   simplified, temperature structures) that, in principle, can be used
   to generate synthesized emission maps for comparison with 
	   the main %martin
	observed PN morphologies across the electromagnetic spectrum. As a first
   step in this direction, we synthesized emission measure distribution
   images (i.e. the integral of $\rho^2$ along the line of sight) for
   Models~1, 2,~3 and 8 using the reconstruction tool
   \textit{Shape} \citep{shape} and following the procedure described
   by \citet{steffen09}.  We choose these models because their shapes
   depend only on their outflow histories. We use logarithmic grayscales
   and an inclination of 90$^{\circ}$ between the polar axis of the
   nebulae and the line of sight (edge-on). Note that in constructing
   these maps, we have not attempted to account for spatial variations
   in gas temperature or ionization state; because much of the gas in
   these simulations is at 
	   time average mean temperatures of order %martin 
	$10^4$\,K, the synthetic images in Figure~7 likely best approximate 
	the appearances the nebulae would have in optical emission lines (e.g., 
	   reflected starlight, emission lines arising in shocks and %from balick and frank 2002
		recombination lines). A more detailed, multiwavelength 
	treatment of the synthesized emission morphologies, which is beyond 
	the scope of this paper, will be the subject of future work.
%%% 
% 

With the above caveats in mind, these initial synthesized emission
maps illustrate the clear relationship between the emission morphology
of a PN and its CSPN outflow history (i.e. collimated jet or isotropic
fast wind). A bipolar object is formed by the interaction of a
dense slow AGB envelope with a brief phase of collimated heavy jet
ejection (\one, Figure~7).
%The emission of the bipolar rim in \one\ shows two bright
%knots along the polar direction which are located symmetrically
%with respect to the center (panel \textit{a} in Figure~6).  Around
%the center we see a fainter emitting region which is disconnected
%form the bright knots. The central region shows sharp horizontal
%edges and resembles a cylinder.
%
The interaction of a light isotropic fast
wind with a spherical AGB envelope yields a spherical nebula (as in \two~~as well). 
The PN rim in these cases shows an emission distribution with a smooth interior surface 
brightness and limb-brightened edges.

%%%%%%%%%%%%%%%%%%%%%%%%%%%%%%%%%%%%%%%%%%%%%%%
\begin{figure*}
\begin{center}
  1 pc \\
  \vskip-.0cm
  \includegraphics[width=.14\textwidth,bb=3.30in 2.0in  4.7in 2.1in,clip=]{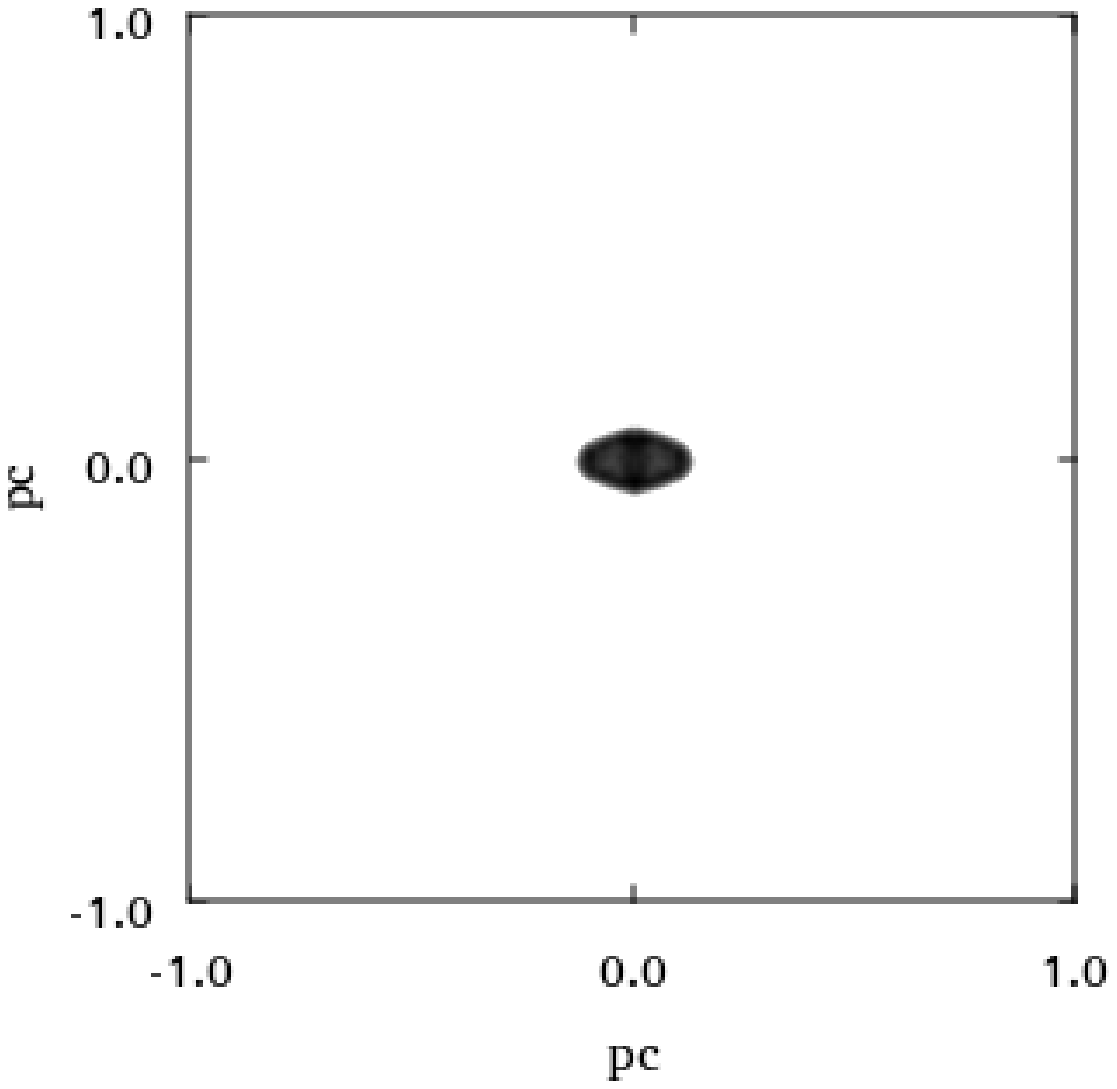}\\
  \end{center}
  \vskip-.0cm
  \one   	   ~~~~~~~~~~~~~~~~~~~~~~~~~~~~~
  \two   				~~~~~~~~~~~~~~~~~~~~~~~~~~~~~~~
  \three 				~~~~~~~~~~~~~~~~~~~~~~~~~~~~
  \eight \\
  \begin{center}
  \vskip-0cm
  \includegraphics[width=.24\textwidth,bb=1.60in 3.5in  5.00in 4.1in,clip=]{jetONLY-cho7-90deg.ps}
  \includegraphics[width=.24\textwidth,bb=1.60in 3.5in  5.00in 4.1in,clip=]{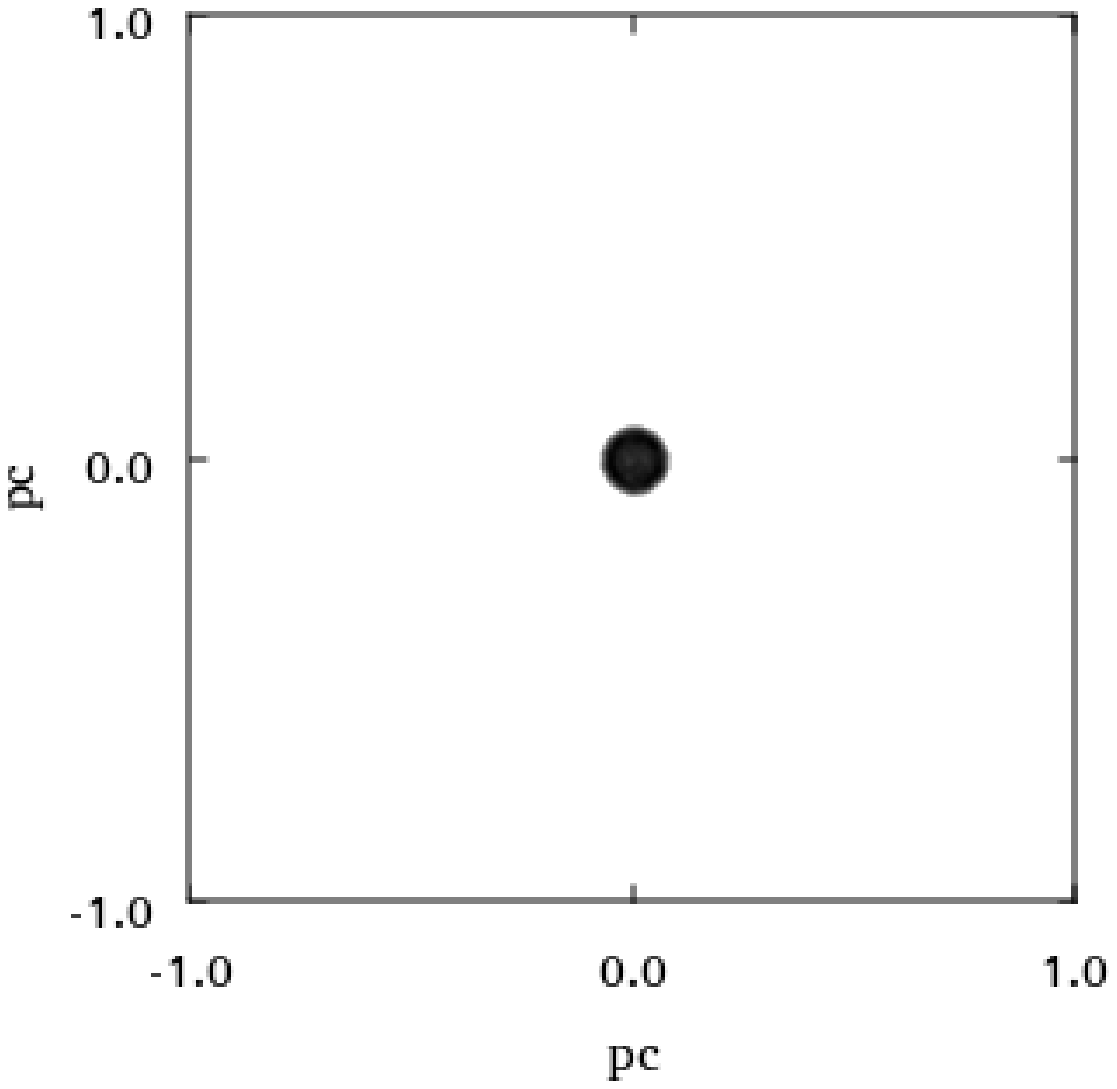}
  \includegraphics[width=.24\textwidth,bb=1.60in 3.5in  5.00in 4.1in,clip=]{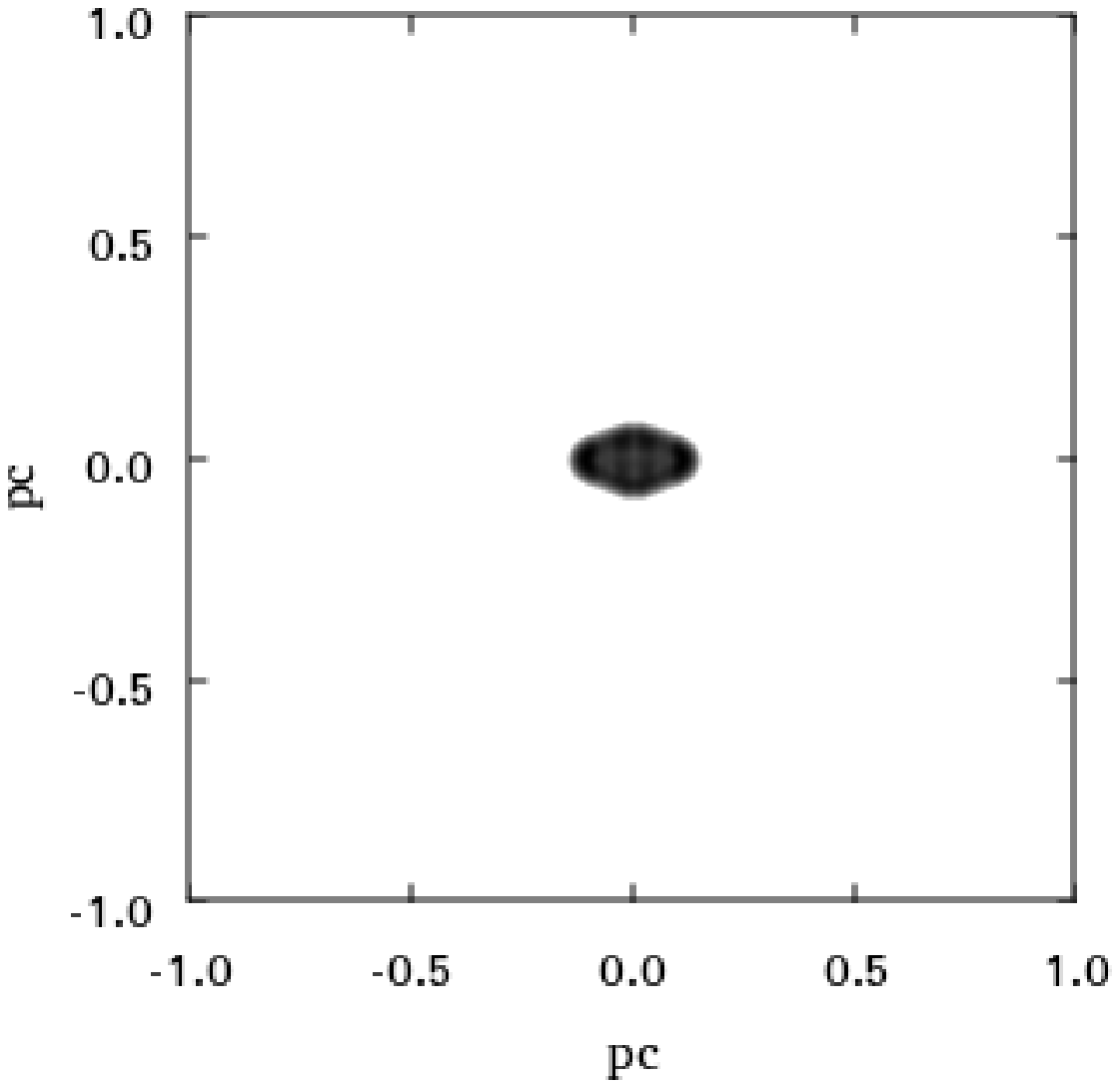}
  \includegraphics[width=.24\textwidth,bb=1.65in 2.95in 5.00in  3.6in,clip=]{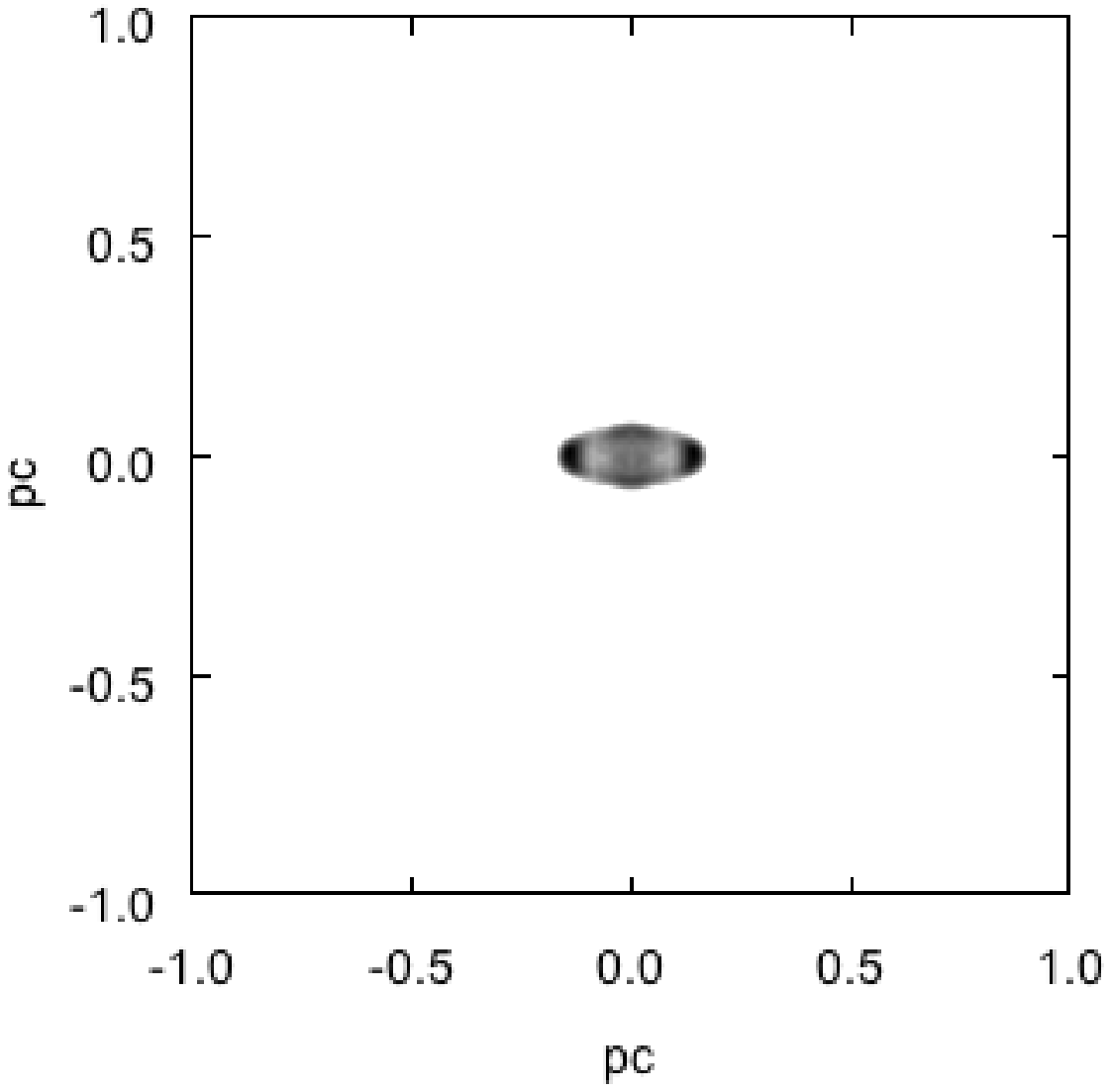}\\
  \vskip-.0cm
  $t \sim \,$700\,yr ~~~~~~~~~~~~~~~~~~~~~~~~~~~~~~
  $t \sim \,$700\,yr ~~~~~~~~~~~~~~~~~~~~~~~~~~~~~~
  $t \sim \,$700\,yr ~~~~~~~~~~~~~~~~~~~~~~~~~~~~~~
  $t \sim\,$1000\,yr\\
  %\one, $t \sim \,$700\,yr. ~~~\two, $t \sim \,$700\,yr. ~~~\three, $t \sim \,$700\,yr. ~~~\eight, $t \sim\,$1000\,yr.\\
  \vskip0.5cm
  {\Large $\downarrow$~~~~~~~~~~~~~~~~~~~~~~~~~~~~~
          $\downarrow$~~~~~~~~~~~~~~~~~~~~~~~~~~~~~
          $\downarrow$~~~~~~~~~~~~~~~~~~~~~~~~~~~~~
          $\downarrow$ }\\
  %
  %\vskip1.5cm
  \includegraphics[width=.24\textwidth,bb=1.6250in  3.00in  4.95in 4.8in,clip=]{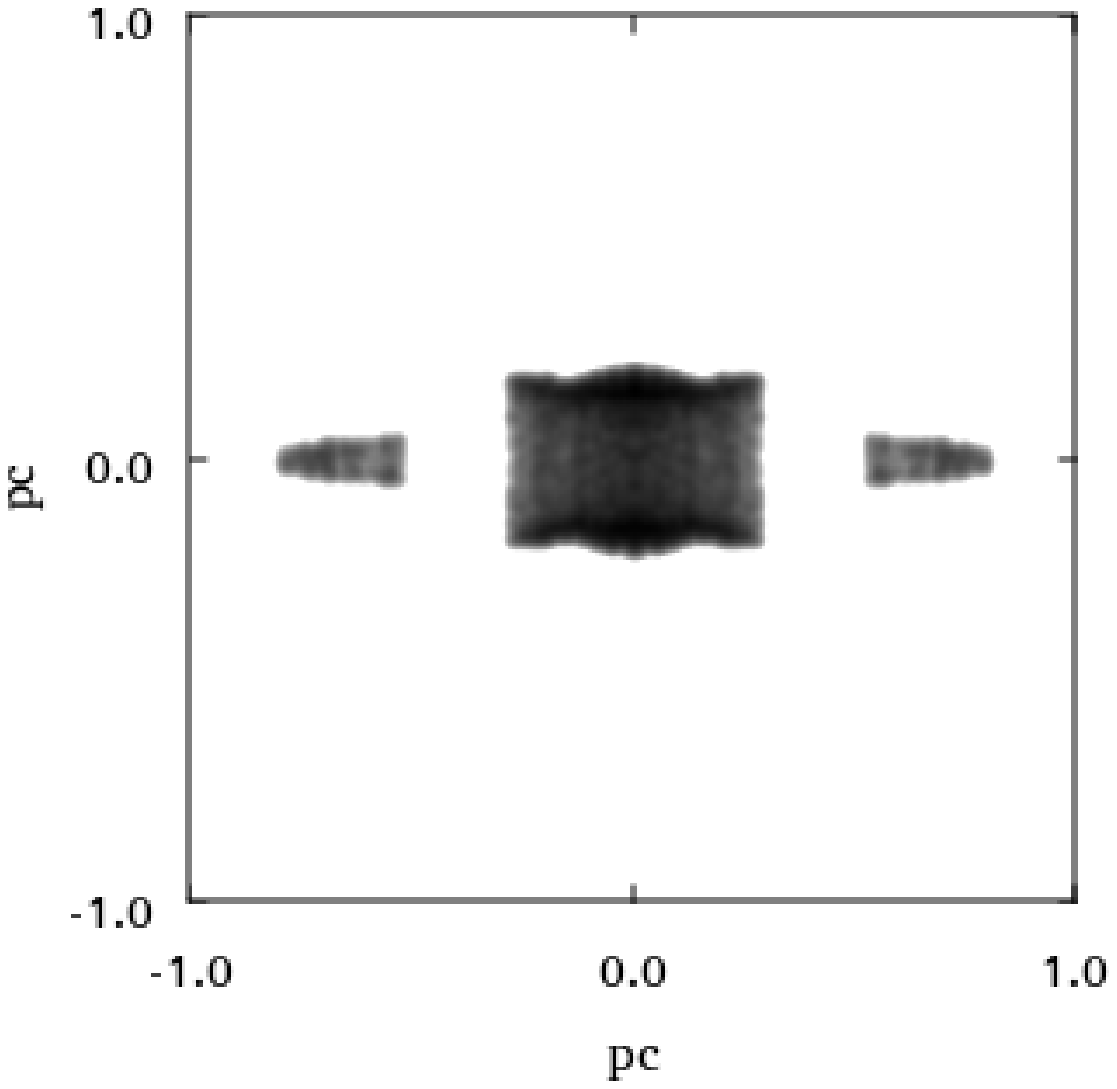}
  \includegraphics[width=.24\textwidth,bb=1.700in  2.40in 4.95in 4.4in,clip=]{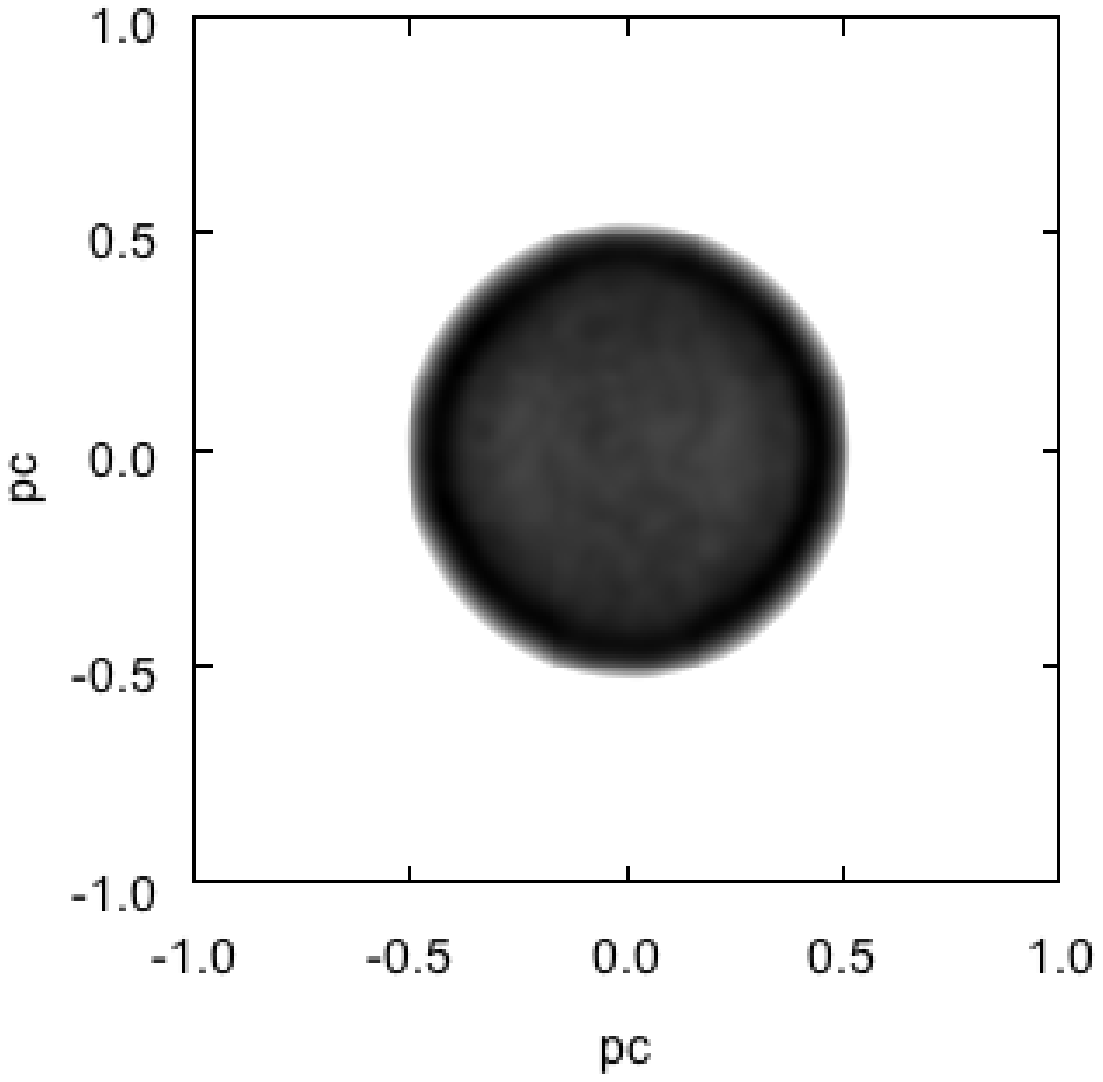}
  \includegraphics[width=.24\textwidth,bb=1.5in  3.00in  4.85in 4.8in,clip=]{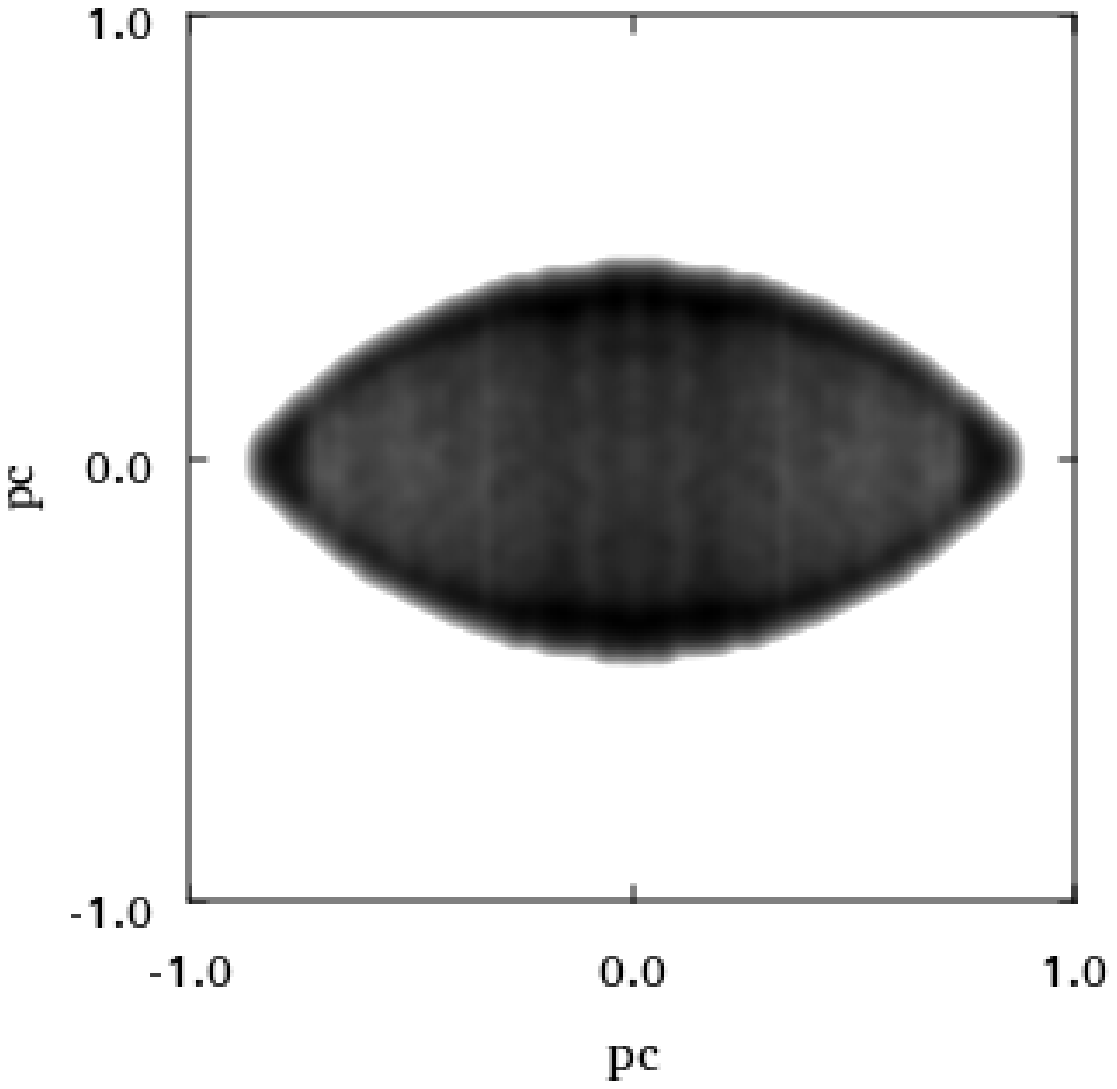}
  \includegraphics[width=.24\textwidth,bb=1.65in  2.50in  5.00in 4.4in,clip=]{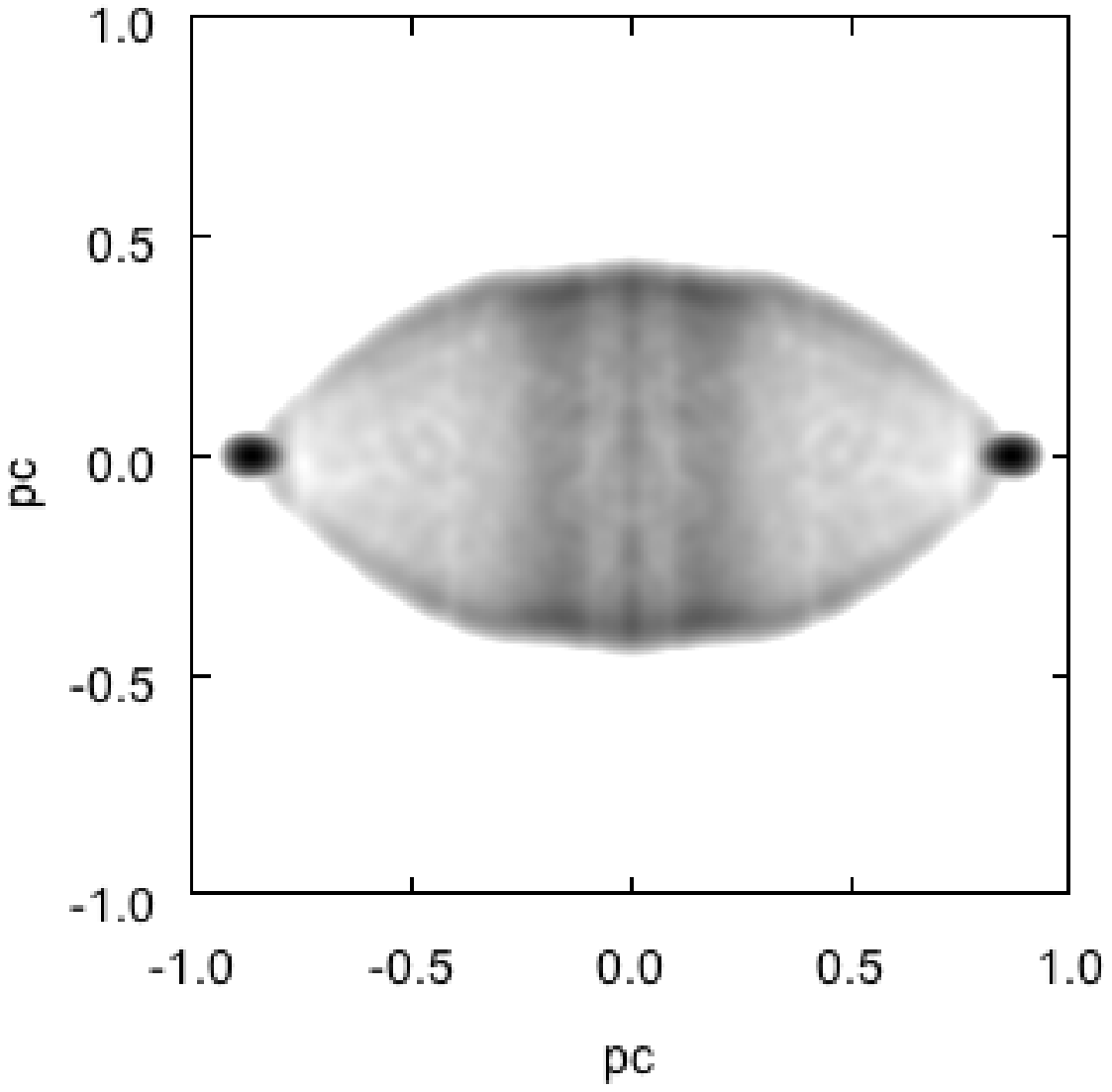}\\
  \vskip-.0cm
  %\one, $t \sim \,$10000\,yr. \two, $t \sim \,$10000\,yr. \three, $t \sim \,$10000\,yr. \eight, $t \sim\,$ 11000\,yr.
  $t \sim \,$10000\,yr ~~~~~~~~~~~~~~~~~~~~~~~~~~
  $t \sim \,$10000\,yr ~~~~~~~~~~~~~~~~~~~~~~~~~~
  $t \sim \,$10000\,yr ~~~~~~~~~~~~~~~~~~~~~~~~~~
  $t \sim\,$ 11000\,yr
  \caption{Synthetic emission maps of the model nebulae. 
  The inclination angle of the images is 90$^{\circ}$. These are
emission measure distribution images (i.e. the integral of $\rho^2$
along the line of sight) produced using Shape
\citep{shape}. Grayscales are logarithmic.
}
     \vspace*{0pt}
\end{center}
\label{synthetic}
\end{figure*}

When all three outflow phases are present, on the other hand, we
see young (not-narrow-waisted) bipolar nebulae (Models~3 and~8, top row, 
Figure~7) transformed 
into older larger elliptical nebulae (bottom row).
The process is particularly evident in \eight\ 
and affirms the behavior seen in the aspect ratio profiles
(Figure~\ref{aspect}).  Such morphological transition 
% 
%%%
   is consistent with the general
  %reproduces the 
%%%
%
trends found by observational studies.

The elliptical rims in cases~3 and~8 have similar
morphologies. However, their synthetic emission during the PN phase is significantly
affected by the presence of the quiescent episode simulated only in
\eight. Figure~7 (bottom) shows that the nebula in \three\ has an emission distribution with
only modest variation in brightness along the rim. In contrast, the PN nebula in
\eight\ shows two bright knots along its polar
axis, located symmetrically with respect to the center.
Examination of the simulation data shows a density contrast of
order~100 between the main part of the nebula and the bright knots.
It is noteworthy that these features bear some morphological
resemblance to the FLIERS that are seen in some elliptical PN
%balick, feb2011 
   (see end of section~\ref{3.1}). 
\subsection{PPN/PN morphological changes and progenitor stars}
\label{stars}

It has been suggested that the outflow mass-loss properties of PN,
in particular their geometry, are affected by both the characteristics
and multiplicity of their progenitor stars
(\citealp{iben91,livio93,soker97}; \citealp[and references
therein]{soker98}; \citealp{nordhaus06,nordhaus07,demarco09,demarco11}).
Where do our models fit in this picture? In other words, which binaries are
expected to produce both spherical AGB envelopes and
jets which could then lead to the morphological change we see in
Models~3 and~8 (from bipolar to elliptical)?

% 
%%%ORSOLA 2 
   Although we have a reasonable understanding of how dust-driven
   winds can explain the high mass-loss rates observed during the
   AGB superwind phase (e.g. Morris 1987; Lagadec \& Zijlstra 2008) 
	questions remain open, in particular regarding
   the geometry change that must take place during the AGB-to-PPN
   transition.
%%% 
% 
%%%ORSOLA: 
	Today, it appears unlikely that a single star may
   power the evolution of an AGB star that results in a non-spherical
   PN (e.g. Nordhaus et al. 2007), although it is possible that in the
   future we may understand how a dynamo could be sustained in an AGB
   star and generate a super-wind phase that departs from spherical
   symmetry (Blackman et al. 2001). The type of binary that could
   give rise to the simulated ejection phases can be surmised from basic
   physical arguments and simulations.

   Companions 
	   that are located %that find themselves %martin 
	within 2-3 AGB star radii can be
   captured into a common envelope (Nordhaus et al. 2010). The subsequent
   in-spiral causes the AGB envelope to be ejected (likely in a toroidal
   geometry; Sandquist et al. 1998) and the post-AGB star to emerge
   with a much reduced radius and much higher temperature. Observed
   systems (e.g., De Marco 2009) and simulations (e.g., Sandquist et
   al. 1998, Ricker and Taam 2008, Passy et al. 2011) indicate that the
   post-common envelope primary will be a star with a radius within a
   solar radius and may bypass the PPN phase altogether, emerging
   primarily as a bipolar PN (although some post-common envelope PN
   are actually elliptical; \citealp{miszalski09a,miszalski09b}).
	It is not clear
   whether a jet would ensue, although highly collimated structures
   are observed in post-common envelope PN (e.g., Abell~63; Mitchell
   et al. 2007). 
	
	% 
	%%% orsola feb'12
	   %Classical common envelope would experience a series
      %of events similar to Models~4, 5~or 6.
      %%%% 
		%%%%%% 
      %%%    Recently, \citet{miszalski11} reported the discovery of ETHOS~1 (PN~G068.1+11.0) which is
		%%%	 a close binary with prominent jets and a seemingly spherical
		%%%	 component to the main nebula. Our models may directly be applied to this system.
		%%%    Other examples of collimated outflows/jets in new
		%%%	 close binaries (which were building on Mitchell et al. 2007) are 
		%%%	 NGC~6326 and NGC~6778 \citep{miszalski11b}, as well as
      %%%    IPHASX~J194359.5+170901A \citep{corradi11}.
		%%%%%% 
		%%%% 
      For some post-CE PN with jets, the jet is kinematically older than
      the main PN and may have started before the bulk of the envelope
      departed the system (e.g. A63, Mitchell et al. 2007) or ETHOS~1
      (Miszalski et al. 2011). For these cases Models 4, 5 or 6 may be
      fitting. In at least one case, however, this is the other way around
      and the jet is younger than the PN  (NGC6~778, Guerrero \& Miranda 2012).
   %%% 
	% 

   A class of binaries exists that has larger final separations
 (100-500~R$_\odot$; \citealp{vanwinckel03,vanwinckel09}).
	It is
   unknown if a common envelope resulted in their formation. Such
   post-AGB primaries are almost never accompanied by a PPN. 
	In a couple of instances bipolar outflows are seen (e.g., the Red
   Rectangle; Cohen et~al. 2004, and references therein). We note that 
	circumbinary disks are observed around all of these system. 
 	These binaries may never develop a proper PN due to their
   envelope being replenished by fall-back of disk material, which
   could prevent a temperature increase of the central star. It is not
   clear if any of our models might mimic this evolutionary channel.

   Other binary types that could fit our models can be wider binaries,
   those with initial separations larger than a few stellar radii
   (5-10~\,AU). These can focus the AGB wind into an equatorially
   concentrated torus \citep{edgar}, accrete matter and blow jets either before
   or after AGB departure. Any of our models could simulate such
   scenario. If the companion is lower mass, both the focussing action and
   the jet blowing may be weak. The star might evolve as if
   single then.

   If a companion is captured tidally by an AGB star {\it and} it has
   considerable mass (larger than a few tenths of a solar mass), it
   will synchronise the AGB stellar rotation with the orbital motion
	\citep{counselman}.
   When this happens the AGB star is spun up and may eject an equatorially
   concentrated wind. A common envelope would be avoided because the
   tidal capture would be slowed down. Jets may be blown for a period
   during or after the AGB envelope ejection. An example of this
   scenario could be Model~6.
   
   Finally, if a companion caught into a common envelope is low mass
   (such as a brown dwarf or even less massive) and tidally destroyed
   within the common envelope, it may form a disk around the core and
   blow jets. In this scenario a jet may precede a possibly toroidal
   AGB wind.
\section{CONCLUSIONS AND SUMMARY} \label{conclu}

We have carried out a series of 2.5D hydro simulations designed to
study how changing the mass-loss properties of PN
progenitor stars affects the evolution of nebular morphology. 
The stellar mass-loss evolution has been followed starting
at the AGB phase, evolving through the PPN phase and finishing at the
mature PN phase.

Our simulations show that a young bipolar PPN transforms into an
older elliptical PN when an initially spherical AGB envelope interacts
with a short duration jet, driven from the central star and  followed by an isotropic fast wind. 

%adam, feb'12:
Thus we have demonstrated how changing
mass-loss geometries from the central engine can drive evolutionary
changes in nebular morphology.  Our theoretical results complement GISW studies.  A
fast spherical CSPN wind driving into a slow spherical AGB wind
has been thought to create a spherical nebula.  We have shown that
if this interaction is preceded by a collimated PPN jet phase then
an elliptical PN will result. A fast spherical CSPN wind driving
into a slow aspherical AGB wind has been thought to create an
aspherical nebula and our results show that including an earlier
PPN jet phase does not affect the eventual butterfly morphology
that results.

Our theoretical result that bipolar PPN can evolve into elliptical PN is
consistent with general observational trends suggesting that PPN appear to favor bipolar morphologies while mature PN appear to be more elliptical.  
This  result  has to be considered along side of  
the complementary  observational trend that   bipolar PN are found
at typically lower Galactic scale heights than ellipticals,  suggesting 
 % progenitor  mass: f
%  i.e., bipolar and ring-like
a correlation between bipolarity and higher-mass (shorter lifetime) progenitors \citep{corradi95,kastner96,manchado11}. 

We also have found that once mature PN become elliptical they do not
evolve to become spherical nebulae over the relevant PN lifetimes, so 
the origin of mature spherical PN (19\% of all PN;
\citealp{parker06,miszalski08}) does not seem to follow the initially aspherical
pathway studied herein.

%MARTIN APR2011 BASED ON SAHAI'S COMMENTS
   The aspect ratio evolution of our model nebulae suggests that
	bipolar PN with projected aspect ratios~$\ga\,$4 may result
	from CSPN which during the post-AGB phase develop a brief jet,
	but no fast spherical winds, or relatively weak ones 
	(e.g., symbiotic Mira systems like R~Aqr, Mz~3, 
	\citealp{Mira}, and references therein;
	Mz~3, \citealp{schwarz92}).
%%% 

When the initial jet drives into a toriodal AGB density distribution
the result is a bipolar PPN that evolves into a mature bipolar nebula
with a narrow waist. The toroidal AGB density distribution in this case
dominates the dynamics and shape of any subsequent outflow phases.
At late times the asymptotic geometry of this nebula is a traditional
butterfly nebula when seen in projection onto the sky.
%the interaction of a dense slow AGB
%envelope with only a brief phase of collimated heavy jet ejection
%forms a bipolar object, the aspect ratio of which increases
%monotonically in time. 
%The interaction of the AGB envelope with
%only a light fast isotropic post-AGB wind yields a spherical PN
%rim.  
%When all three phases are modeled, however, we find the
%bipolar cavity driven by the jets is transformed into older
%larger elliptical nebulae that expands homologously with a projected
%aspect ratio close to~2. 
%Thus if the generic sequence of mass loss
%evolution is a spherical AGB, followed by a collimated PPN jet,
%followed by a spherical PN wind, then the fluid dynamics in our
%simulations provide a means for accounting for the differing
%statistics of PPN and PN morphologies.
%
%The shape of our model nebulae also correlates well
%with the distribution of both the density and the velocity of the
%AGB envelope as has been seen in previous simulations.  
%Narrow-waisted butterfly shaped objects are also formed 
%for this outflow phase sequence which is dominated by the
%toroidal density distribution of the AGB envelope.
%by the interaction
%of any post-AGB outflows and
%a toroidal AGB. %envelope with a toroidal density distribution. 
%The aspect 
%ratio of these objects increases rapidly in time, and if a fast
%wind episode is present at late times, the aspect ratios reach a
%maximum of about~3, when the objects are $\sim\,$0.4\,pc long, and
%decrease slowly afterwards, down to $\sim\,$1.8.  
We also find that the interaction of a spherical
post-AGB fast wind with an AGB envelope which has an aspherical
velocity field yields a homologously expanding elliptical nebula.
%The synthetic emission distribution of this nebulae, moreover, shows
%anisotropies which, in comparison with the smooth synthetic emission
%of a spherical nebula (formed by the interaction of the fast wind
%and a spherical AGB envelope), are possibly related to the aspherical
%velocity field of the AGB envelope. We have found qualitative
%similarities between our synthetic emission maps and some PN
%observations.
%
%The evolution of these rims, however, does not reproduce the
%observed morphological histories of PPN and PN.

One of our simulations included a quiescent interlude between the jet
and the fast wind episodes. We find the corresponding synthetic
emission map shows small bright features which resemble observations
of axial knots or FLIERS within PN. A future avenue for these studies
would be to explore when and how knots form as the result of
jet material collecting at the head of the flow.
%balick
   %While we did not find
   %a rapid temperature increase due to ionization changed our
   %results, future studies should include the effect of ionization.
%%

\section*{Acknowledgments}
Financial support for this project was provided by the Space Telescope
Science Institute grants HST-AR-11251.01-A and HST-AR-12128.01-A;
by the National Science Foundation under award AST-0807363; by the
Department of Energy under award DE-SC0001063 and by Cornell
University grant 41843-7012. 
% 
%%%JOEL 
JHK acknowledges financial support by award number GO1-12025A issued
to RIT by the Chandra X-ray Observatory Center, which is operated
by the Smithsonian Astrophysical Observatory for and on behalf of
NASA under contract NAS8-03060. 
%%% 
% 
MHE thanks Jonathan Carroll for discussions.
We thank the referee, Wolfgang Steffen,
for useful comments that helped to improve this paper.

\bsp

\label{lastpage}
 
\end{document}